\documentclass[12pt]{article}
\usepackage{geometry}
\usepackage{authblk} 
\usepackage{lineno} 
\usepackage{microtype}
\usepackage{amsbsy} 
\usepackage{amssymb}
\usepackage{amsmath}
\usepackage[round]{natbib}
\usepackage{csvsimple}
\usepackage{footnote}
\usepackage{tablefootnote}
\usepackage{array}
\usepackage{booktabs}
\usepackage{adjustbox}
\usepackage{multirow}
\usepackage{arydshln}
\usepackage{pdfpages}
\usepackage{graphicx}
\usepackage{caption}
\usepackage[justification=justified,singlelinecheck=false]{subcaption}
\usepackage{setspace}

\begin{document}

\title{A Non-ergodic Spectral Acceleration Ground Motion Model for California Developed with Random Vibration Theory 
       [Submitted to Bulletin of  Earthquake Engineering]}

\author[1]{Grigorios Lavrentiadis\thanks{glavrent@berkeley.edu}}
\author[1]{Norman A. Abrahamson\thanks{abrahamson@berkeley.edu }}
\affil[1]{Department of Civil Engineering, University of California, Berkeley}

\renewcommand\Authands{ and }

\maketitle

\begin{abstract}

A new approach for creating a non-ergodic $PSA$ ground-motion model (GMM) is presented which account for the magnitude dependence of the non-ergodic effects.
In this approach, the average $PSA$ scaling is controlled by an ergodic $PSA$ GMM, and the non-ergodic effects are captured with non-ergodic $PSA$ factors, which are the adjustment that needs to be applied to an ergodic $PSA$ GMM to incorporate the non-ergodic effects. 
The non-ergodic $PSA$ factors are based on $EAS$ non-ergodic effects and are converted to $PSA$ through Random Vibration Theory (RVT).
The advantage of this approach is that it better captures the non-ergodic source, path, and site effects through the small magnitude earthquakes.
Due to the linear properties of Fourier Transform, the $EAS$ non-ergodic effects of the small events can be applied directly to the large magnitude events.
This is not the case for $PSA$, as response spectrum is controlled by a range of frequencies, making $PSA$ non-ergodic effects depended on the spectral shape which is magnitude dependent. 

Two $PSA$ non-ergodic GMMs are derived using the ASK14 \citep{Abrahamson2014} and CY14 \citep{ChiouYoungs2014} GMMs as backbone models, respectively.
The non-ergodic $EAS$ effects are estimated with the LAK21 \citep{Lavrentiadis2021} GMM. 
The RVT calculations are performed with the V75 \citep{Vanmarcke1975} peak factor model, the $D_{a0.05-0.85}$ estimate of AS96 \citep{Abrahamson1996} for the ground-motion duration, and BT15 \citep{Boore2015} oscillator-duration model. 
The California subset of the NGAWest2 database \citep{Ancheta2014} is used for both models. 

The total aleatory standard deviation of the two non-ergodic $PSA$ GMMs is approximately $30$ to $35\%$ smaller than the total aleatory standard deviation of the corresponding ergodic $PSA$ GMMs. 
This reduction has a significant impact on hazard calculations at large return periods. 
In remote areas, far from stations and past events, the reduction of aleatory variability is accompanied by an increase of epistemic uncertainty. 


\end{abstract}

\section{Introduction} \label{sec:intro}
\begin{figure}
    \centering
    \includegraphics[width=0.5\textwidth]{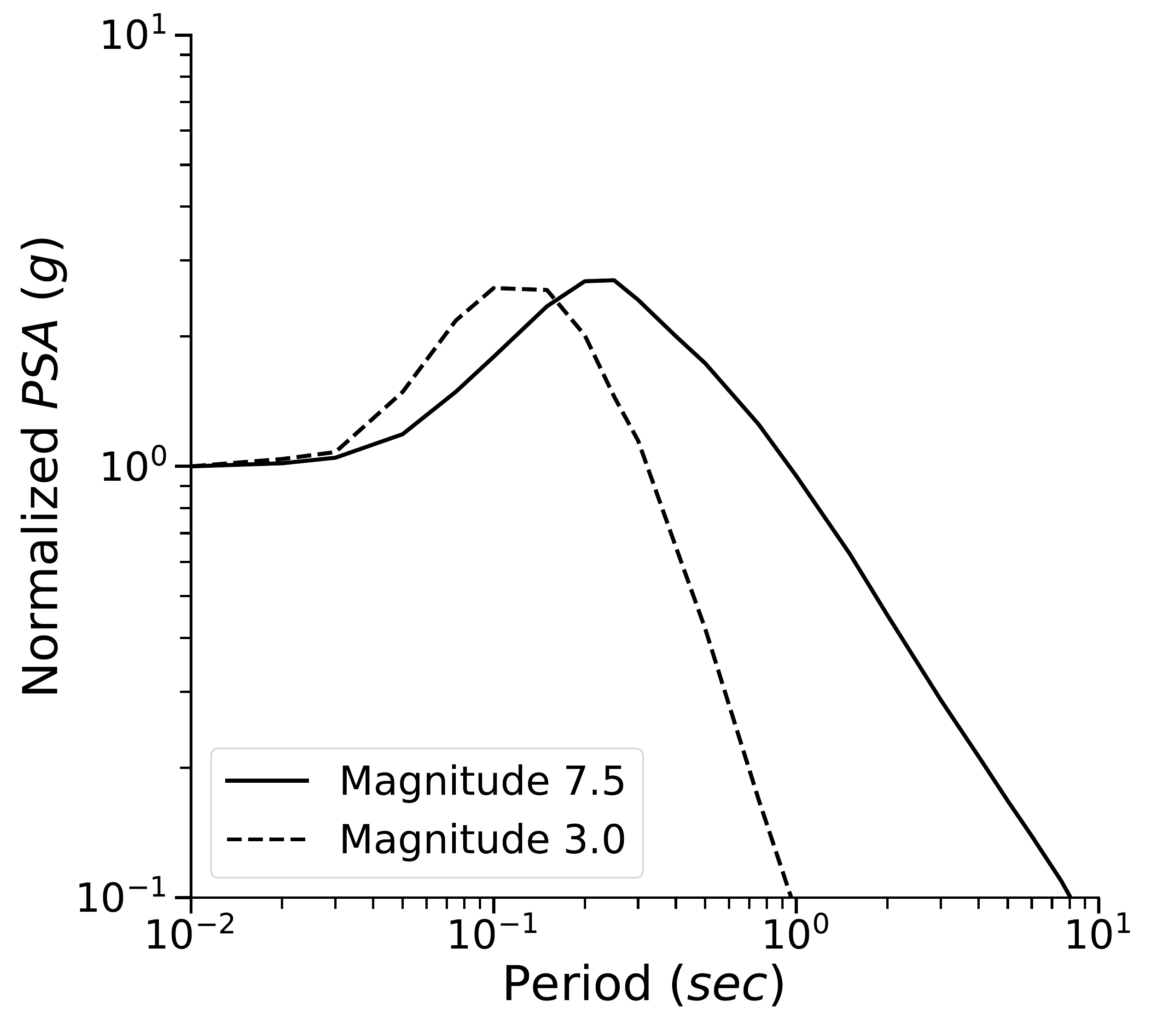}
    \caption{Schematic of normalized response spectra for $M$ $3.0$ and $7.5$ earthquakes}
    \label{fig:sketch_psa}
\end{figure}

Ground-motion models (GMMs) are used to estimate the distribution of a ground-motion intensity measure ($IM$) for a given earthquake scenario. 
The most common IM is pseudo-spectral acceleration ($PSA$) as it is a good estimator of seismic loading for a wide range of structures. 
$PSA$ is defined as the absolute maximum response of a single-degree-of-freedom oscillator (SDOF) to an input ground motion.
SDOFs are defined by their natural period ($T_0$) or natural frequency ($f_0 = 1/T_0$) and damping ($\zeta$); in GMMs, typically, $T_0$ ranges from $0.01$ to $10~sec$ and, $\zeta$ is equal to $5 \%$.
The response of the oscillator depends on the frequency content and timing (compactness of energy) of the ground motion. 
From the entire frequency content of the ground motion, the response of the oscillator mainly depends on the amplitudes of the frequencies near and below  $f_0$.
Therefore, at small $T_0$ (high $f_0$), the response of the oscillator depends on the entire frequency content of the ground motion (i.e. spectral shape) and not just a narrow frequency bin.
This makes the coefficients of a $PSA$ GMM at small $T_0$ magnitude dependent even for linear effects, as the shape of spectral acceleration response spectrum changes with magnitudes.
The peak of a spectral acceleration response spectrum will be at $0.1~sec$ for a magnitude ($M$) $3$ event and at $0.3~sec$ for a $M~7.5$ event (Figure \ref{fig:sketch_psa}); this means that at small magnitudes, the $PGA$ scaling (e.g. $V_{S30}$ coefficient) will be consistent with the scaling of $T_0=0.1~sec$, while at large magnitudes, the $PGA$ scaling will be consistent with $T=0.3sec$.
This is also observed by \cite{Stafford2017}, who showed that the linear site amplification factors are magnitude and distance dependent.
A detailed discussion the differences between the scaling of $FAS$ and $PSA$ is given by \cite{Bora2016}.

Most $PSA$ GMMs do not explicitly account for the magnitude dependence of the coefficients, such as the $V_{S30}$ scaling or distance scaling; instead, they often use a limited range of magnitudes where the magnitude dependence of the coefficients is not pronounced. 
For instance, the data-set that was used in the development of the NGA West1 GMMs had a limited set of magnitudes that ranged from $M~4.5$ to $M~8$ \cite{Power2008}.
The approach of using a smaller range of magnitudes works when developing an ergodic GMM, as there is enough number of moderate-to-large magnitude events globally to estimate the coefficients, but it can be problematic when developing a non-ergodic GMM. 

For the NGA West2 GMMs, the data set was extended to down to $M 3$ with the objective of setting the reference ergodic model that could be used to evaluate regional differences in the site, path, and source terms based on small magnitude data.
The NGA West2 GMMs modified the magnitude scaling to capture the average effect of the magnitude dependence of the coefficients, but this does not accurately model the magnitude dependence of the site and path effects. 

GMMs fall into two main categories: ergodic GMM and non-ergodic GMM.
Ergodic GMMs assume that the statistical properties of a ground motion IM do not change in space \citep{Anderson1999}, and therefore, earthquakes and recordings from all around the world can be merged into a single dataset to estimate the GMM coefficients. 
Models developed under this assumption tend to have stable median estimates but large aleatory variability.
Some models developed with the ergodic approach are: the NGA West GMMs for California \cite{Abrahamson2008}, and the \cite{Douglas2014} GMM for Europe.
Non-ergodic GMMs recognize that source, path, and site effects are systematically different at different parts of the world and account for these differences in the model development.
Non-ergodic GMMs have smaller aleatory variability than ergodic GMMs, but in areas with sparse data, where the systematic effects are unknown, the reduced aleatory variability is accompanied by an increase in the epistemic uncertainty of the values of the median ground motion. 
The use of non-ergodic GMMs in Probabilistic Seismic Hazard Analysis (PSHA) is very promising, as the reduction in aleatory variability can have a large impact on the seismic hazard at large return periods, improving the accuracy of the site-specific hazard.
A more in-depth discussion of ergodic and non-ergodic GMM is provided in the accompanying paper \cite{Lavrentiadis2021}.

The estimation of the non-ergodic terms requires a large set of regional data.
To achieve that, the datasets used in the development of non-ergodic GMM need to have a wider range of magnitudes to include the more frequent small-to-moderate earthquakes.
It is this expansion of the magnitude range that makes the magnitude dependence of the GMM coefficients a more significant issue in non-ergodic GMMs. 
One solution to this problem is, first, develop a non-ergodic GMM for an $IM$ whose scaling does not suffer from the magnitude dependence, as $PSA$ does, and then for a scenario of interest, calculate the non-ergodic $PSA$ based on the non-ergodic $IM$ estimate. 

The effective amplitude spectrum ($EAS$), defined in \cite{Goulet2018}, is one such $IM$: the $EAS$ is a smoothed rotation-independent average power Fourier amplitude spectrum ($FAS$) of the two horizontal components of an acceleration time history. 
In $EAS$, the amplitude at each frequency is independent of the amplitudes of the adjacent frequencies making the coefficients of an $EAS$ GMM magnitude independent.
Random vibration theory (RVT) provides a framework to calculate $PSA$ from $EAS$.
It relies on extreme-value statistics to estimate the peak response of the oscillator directly in the Fourier domain; it does not require a phase-angle spectrum to first convert the ground motion in the time domain to compute the peak oscillator response.
RVT has been used in the past to compute $PSA$ based on $FAS$ from seismological theory \citep{Hanks1981, Boore1983, Boore2003}
Other studies, such as \cite{Boore1984}, \cite{Liu1999} \cite{Bora2015} and, \cite{Boore2012}, focused on semi-empirical adjustments to the RVT framework to correct for the assumptions not satisfied by ground motions, mainly the fact that acceleration time histories are not stationary signals.
More recently, \cite{Kottke2021} used RVT to develop an ergodic $PSA$ GMM for the eastern US based on an ergodic $EAS$ GMM for the same region. 

In this study, we developed two non-ergodic $PSA$ GMM.
The average $PSA$ scaling is determined by backbone ergodic $PSA$ GMMs.
The non-ergodic effects are defined in terms of non-ergodic $PSA$ factors which are estimated by combining the \cite{Lavrentiadis2021} non-ergodic $EAS$ GMM with RVT.

\section{Ground-Motion Data} \label{sec:data} 

\begin{figure}
    \centering
    \includegraphics[width = 0.5 \textwidth]{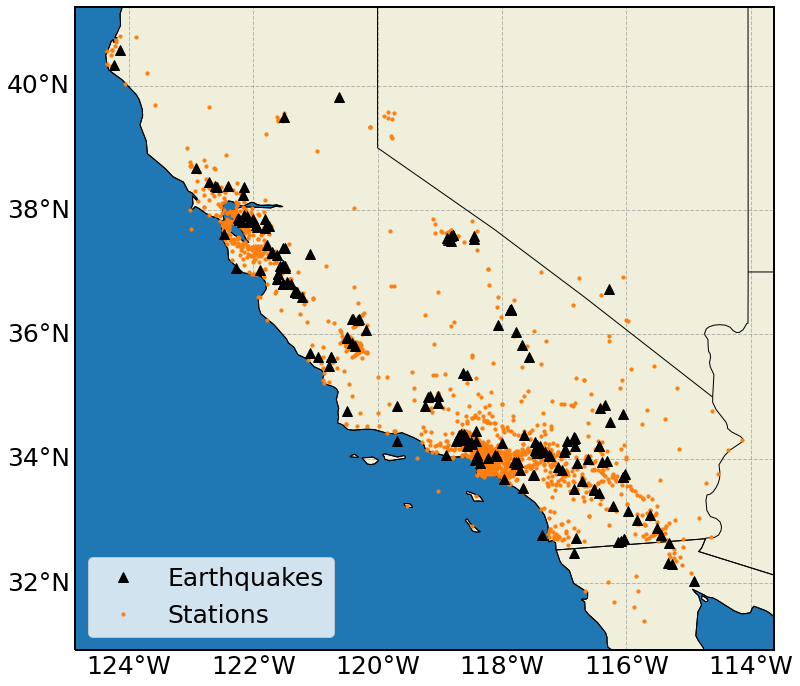}
    \caption{Spatial distribution for earthquakes and station used in this study.  }
    \label{fig:dataset_spat_dist}
\end{figure}

\begin{figure}
    \centering
    \begin{subfigure}[t]{0.40\textwidth} 
        \caption{} 
        \includegraphics[width=.95\textwidth]{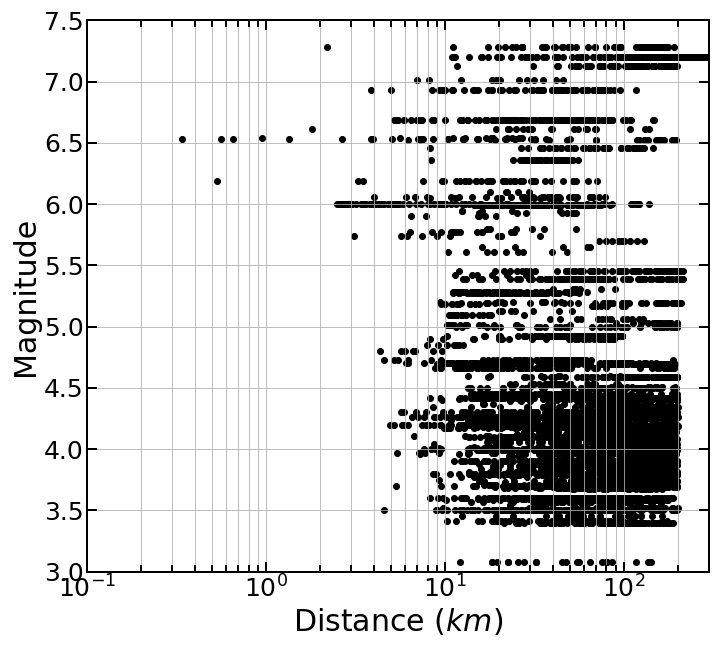}
    \end{subfigure}
    \begin{subfigure}[t]{0.40\textwidth}
        \caption{}  
        \includegraphics[width=.95\textwidth]{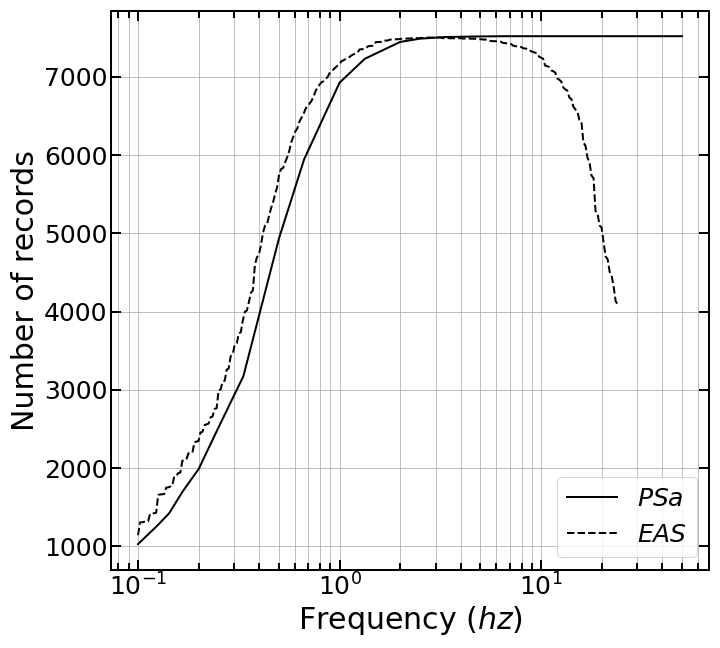}
    \end{subfigure}
    \caption{Selected data from the NGAWest2 database. 
    (a) Magnitude - Distance distribution, 
    (b) number of $PSA$ and $EAS$ recordings per frequency used in the regression analysis}
    \label{fig:dataset_nga2CA}
\end{figure}

A subset of the NGAWest2 data-set \citep{Ancheta2014} was used in this study. 
The selected subset contains the earthquake and stations that are located in California, western Nevada, and northern Mexico. 
Recordings that were flagged as questionable in \cite{Abrahamson2014} were removed from the regression subset
Figure \ref{fig:dataset_spat_dist} shows the spatial distribution of earthquakes and stations.
Most of the stations are located in Los Angeles, Bay Area, and San Diego metropolitan areas, whereas spatial density of the stations is lower in less populated areas, such as northern-eastern California.
The regression data-set contains $7520$ records from $185$ earthquakes recorded at $1410$ stations.
Figure \ref{fig:dataset_nga2CA} shows the magnitude-distance distribution of the data and the number of records per frequency.
The magnitude of the earthquakes ranges from $3.1$ to $7.3$, and the distance of most records ranges from $10$ to $200~km$. 
The usable frequency range of the majority of $EAS$ records spans from $0.4$ and $20 Hz$.
The minimum usable frequency of most $PSA$ records is $0.5~Hz$.

\section{Model development} \label{sec:mod_dev}

\subsection{Random-Vibration Theory} \label{sec:rvt}

RVT uses Parseval's theorem and extreme-value statistics ($EVS$) to estimate the $PSA$ based on the frequency content (i.e. $FAS$) and duration of a ground motion.
Parseval's theorem is used to calculate the root-mean-square of the oscillator's response ($x_{rms}$) to the input ground motion, and a peak factor ($PF$), based on $EVS$, is used to estimate the absolute peak response of the oscillator, which is the definition of $PSA$, based on $x_{rms}$. 
$PFs$ assume that the ground motion is a stationary stochastic process, and that it can be described as a band-limited white Gaussian noise with zero mean.
The first assumption means that the amplitudes of the ground motion are identically distributed, and the second assumption means that the phase angles of the ground motion are randomly distributed.
Although, earthquake ground motions violate both assumptions, numerous studies have shown that RVT provides $PSA$ estimates that are in agreement with observed ground motions \citep{Hanks1981, Boore1983, Boore2003}

\subsubsection{Oscillator Response} \label{sec:sdof}

\begin{figure}
    \centering
    \begin{subfigure}[t]{0.40\textwidth} 
        \caption{} 
        \includegraphics[width=.95\textwidth]{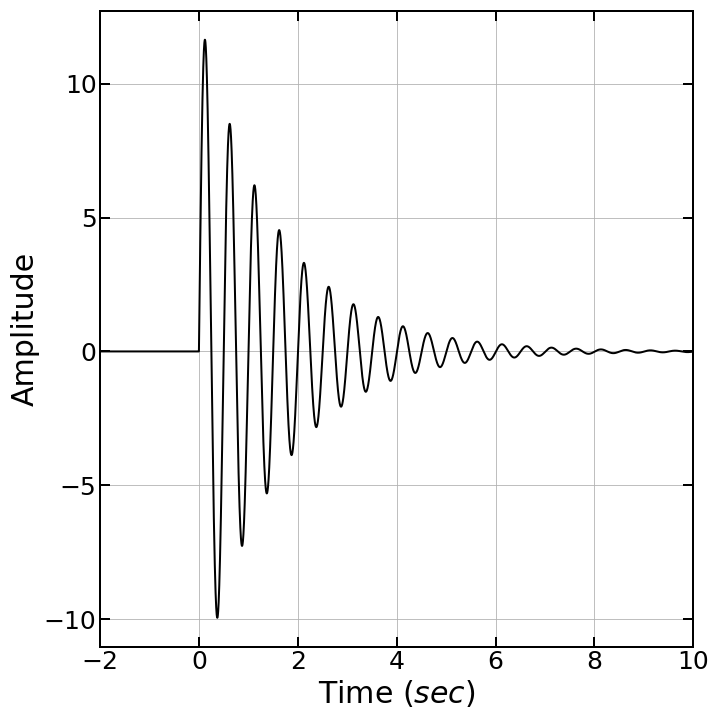}
    \end{subfigure}
    \begin{subfigure}[t]{0.40\textwidth}
        \caption{}  
        \includegraphics[width=.95\textwidth]{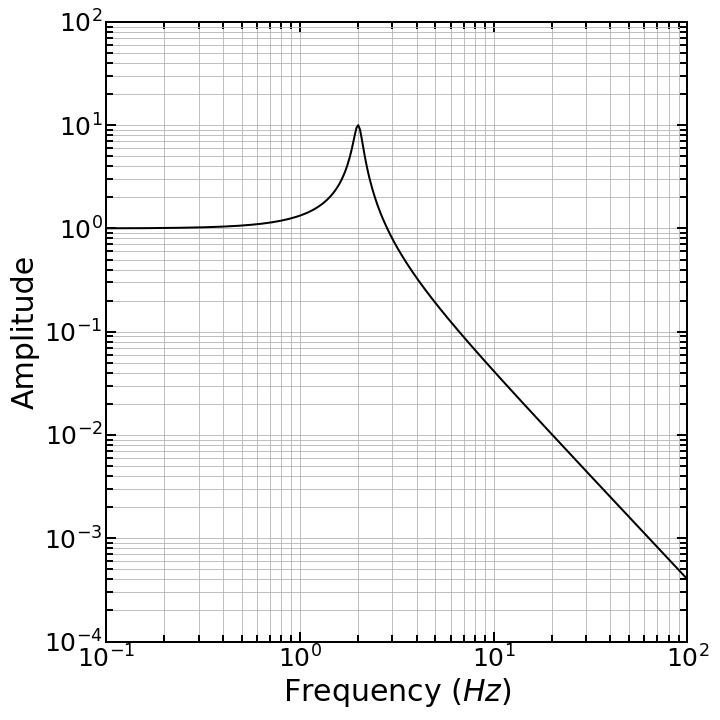}
    \end{subfigure}
    \caption{Impulse response of a single degree of oscillator;
    (a) Time domain, 
    (b) Fourier domain}
    \label{fig:imp_response}
\end{figure}

The response of an oscillator to a ground motion can be computed by convolving the ground motion with the impulse response ($IR$) of the oscillator.
The $IR$ is the response of an oscillator to a very brief acceleration pulse; that is a Dirac delta function. 
For an SDOF oscillator, the Fourier transform of the impulse response is:

\begin{equation} \label{eq:IR}
    IR(f, f_0,\zeta) = \frac{- f^2_0}{f^2 - f_0^2 - 2j * \zeta * f_0 * f}
\end{equation}

\noindent where, $f_0$ is the natural frequency of the oscillator, and $\zeta$ is the damping of the oscillator.
As an example, Figure \ref{fig:imp_response} shows the $PSA$ impulse response, in time and Fourier domain, for an SDOF oscillator with $f_0=2Hz$ and $\zeta=5\%$. 
In the Fourier domain, the convolution is performed by multiplying the ground motion's $FAS$ with $IR$; therefore, the response of an SDOF oscillator to a ground motion is:

\begin{equation} \label{eq:convFT}
    X(f) = FAS(f) ~ IR_{SD}(f, f_0,\zeta)
\end{equation}

The $x_{rms}$ of the oscillator's response is defined as:

\begin{equation} \label{eq:rms}
    x_{rms} = \sqrt{ \frac{1}{D_{rms}} \int_{-\infty}^{+\infty} x(t)^2 dt  }
\end{equation}

\noindent where $D_{rms}$ is a measure of the duration which is defined in Section \ref{sec:dur_non_stat}.
Parseval's theorem states that the amount of energy in the time domain is equal to the amount of energy in the Fourier domain ($\int_{-\infty}^{+\infty} x(t)^2 dt = 2 \int_{0}^{+\infty} X(f)^2 df$) which allows to compute $x_{rms}$ directly in Fourier domain:

\begin{equation} \label{eq:rmsFT}
    x_{rms} = \sqrt{ \frac{1}{D_{rms}} 2 \int_{0}^{+\infty} X(f)^2 df  } = \sqrt{ \frac{m_0}{D_{rms}} } 
\end{equation}

\noindent with $m_0$ being the zeroth moment of $FAS$. The $k^{th}$ moment of $FAS$ is defined as:

\begin{equation} \label{eq:moments}
    m_{k} = 2 \int_{0}^{+\infty} (2 \pi f)^k X(f)^2 df   
\end{equation}

\subsubsection{Peak Factor}

The peak factor relates the $x_{rms}$ with the maximum response of the oscillator ($x_{max}$), which is the definition of the $PSA$.

\begin{equation} \label{eq:PF}
    PSA = PF ~ x_{rms}
\end{equation}

In general, $PFs$ fall into two main categories: those based on the \cite{Cartwright1956} peak factor, abbreviated as CLH56, and those that are based on the \cite{Vanmarcke1975} peak factor, abbreviated as V75.

In the first group, the CLH56 peak factor assumed that the peaks of a time history occur independently according to a Poisson process.
In a series of papers, Boore and colleagues  \citep{Boore1983, Boore1984, Boore2003} developed peak factors (BJ83) based on a reformulated version of CLH56 and removed an integrable singularity.
\cite{Davenport1964} proposed the a peak factor model (D64) based on an asymptotic form that approximates CLH56 for long time histories. 

The main difference between V75 \citep{Vanmarcke1975, Vanmarcke1976} and the $PFs$ of the first group is that V75 dropped the Poisson process assumption.
Because of this, V75 $PF$ accounts for the time spend outside the threshold, which is important for a narrow-band process, and considers that the peaks could be clustered in time, which is important for a wide-band process. 
\cite{DerKiureghian1980} noted that the D64 peak factor overestimates the number of zero crossings, and developed a  new $PF$ model (DK80) by modifying D65 $PF$ so that it is asymptotically consistent with V75. 
V75 and D80 are in general agreement, but they deviate in time histories with a small number of zero crossings. 

The V75 $PF$ is selected for the development of the non-ergodic $PSA$ GMM.
V75 is preferred over the group of $PF$ that are based on CLH56 due to the simplified assumptions in CLH56, and the complete form of V75 is preferred over the asymptotic forms, as the former is more accurate for the wide range of ground motions considered in this project. 
This choice is consistent with the $PF$ used in \cite{Kottke2021}. 

V75 expressed the probability distribution of the peaks as a first-passage problem. 
For a Gaussian process, the first-passage probability (i.e. the probability of no crossing) a $\pm a$ threshold (type-D barrier) in the time interval $(0,t)$ is equal to:


\begin{equation} \label{eq:V75_FP}
    P( |z| < r) = A ~ \exp\left( - f_z t ~ \exp(-r^2/2)  \frac{ 1-\exp(-\sqrt{\pi/2} ~ \delta_e ~ r) }{ 1 - \exp(-r^2/2)} \right)
\end{equation}

\noindent where $r$ is the normalized barrier level ($r=a/x_{rms}$), $A$ is the probability of starting within the thresholds ($A = 1-\exp(-r^2/2)$), $f_z$ is the average rate of zero crossings, and $\delta_e$ is an semi-empirical measure of bandwidth ($\delta_e = \delta^{1+b}$). 
$b$ a non-negative constant which, in this case, is equal to $0.2$, and $\delta$ is a measure of bandwidth based on the spectral moments \citep{Vanmarcke1972} defined as:

\begin{equation}
    \delta = \sqrt{1- \frac{m^2_1}{m_0 m_2}}
\end{equation}

The cumulative distribution function (CDF) of the peak values is obtained by setting $t$ equal to $D_{gm}$ in equation \eqref{eq:V75_FP}; that is, the probability of the peak of the time history being less than $r \times x_{rms}$ is equal to the probability that the time history will remain within the thresholds $\pm r \times x_{rms}$ for the entire ground-motion duration. 
With that, the CDF of $PF$ is equal to:

\begin{equation} \label{eq:V75_PF}
\begin{aligned}
    F_{PF}( r ) =& \left( 1-\exp(-r^2/2) \right) \\
                 &\times \exp\left( - f_z D_{gm} ~ \exp(-r^2/2)  \frac{ 1-\exp(-\sqrt{\pi/2} ~ \delta_e ~ r) }{ 1 - \exp(-r^2/2)} \right)    
\end{aligned}
\end{equation}

The expected value of $PF$ can be computed with the probability density function (PDF) of $PF$ (Equation \eqref{eq:PF_mean_pdf}), which requires the derivation of the PDF.
However, $PF$ is continuous and defined on the positive side of the real line; thus, the expected value of $PF$ can be computed directly from the CDF with equation Equation \eqref{eq:PF_mean_cdf}.

\begin{equation} \label{eq:PF_mean_pdf}
    E[PF] = \int_0^{+\infty} r f_{PF}(r) ~ dr
\end{equation}

\begin{equation} \label{eq:PF_mean_cdf}
    E[PF] = \int_0^{+\infty} \left( 1- F_{PF}(r) \right) ~ dr
\end{equation}
 
The mean estimate of the RVT $PSA$ can be computed by substituting the expected value of the $V75$ $PF$ in Equation \eqref{eq:PF}. 

\subsubsection{Ground-Motion Duration} 

In RVT, a measure of duration is needed in two steps: in the calculation of the peak factor, and in the calculation of $x_{rms}$.
Due to transient nature of a ground-motion, the duration measures used in these two steps are often different. 
$D_{gm}$ is the ground-motion duration, which is used in the calculation of $PF$; $D_{rms}$ is the duration measure for the calculation of $x_{rms}$. which is defined in section \ref{sec:dur_non_stat}.


In seismology, the ground-motion duration is most commonly defined either as bracketed or as significant duration.
Bracketed duration is the time interval between the first and last time the ground motion exceeds a threshold.
Significant duration is the difference in time the normalized Arias intensity reaches two specific values. 
For instance, the $5 - 75 \%$ significant duration is the difference between the time the normalized Arias intensity is $5 \%$ and, the time the normalized Arias intensity is $75 \%$. 
The Arias intensity is defined as integral of the squared acceleration time history:

\begin{equation}
    I_a(t') = \frac{2\pi}{g} \int_0^{t'} x^2(t) dt
\end{equation}

The normalized Arias intensity, also known as Husid curve, is the ratio of $I_a$ at time $t$ over $I_a$ at the end of the ground motion:

\begin{equation} 
    h(t) = \frac{I_a(t)}{I_a(+\infty)} 
\end{equation}

In some RVT methods, $D_{gm}$ is set to a measure of significant duration, but in others, $D_{gm}$ is treated as a free parameter with units of time.  
For instance, \cite{Boore2003} used the $D_{a 0.05-0.95}$ significant duration as $D_{gm}$, while \cite{Bora2015} and \cite{Bora2019} treated $D_{gm}$ as free parameter and developed a duration GMM with the goal to minimize misfit between the observed $PSA$ and the $PSA$ computed with RVT. 

In this study, $D_{gm}$ is defined as an interval of significant duration. 
Different intervals of significant duration were tested as $D_{gm}$ candidates to find the one that minimized the misfit between the $PSA$ of the used dataset ($PSA_{NGA}$) and the $PSA$ estimated with RVT ($PSA_{RVT}$); the results of this comparison are shown in the Electronic supplement, Section S1. 
The $D_{a0.05-0.85}$ significant duration resulted in the best fit of $PSA_{NGA}$ for the entire frequency range, $0.1$ to $100~Hz$. 
The \cite{Abrahamson1996} duration GMM (AS96) was selected for estimating $D_{a0.05-0.85}$ for new scenarios, as to our knowledge, AS96 is the only GMM that provides an estimate for the selected duration interval. 
Despite the previous results, the $D_{a5-75}$, $D_{a5-95}$, $D_{v5-75}$, and $D_{v5-95}$ estimates of the \cite{Kempton2006} duration GMM and $D_{a5-75}$, $D_{a5-95}$, and $2D_{a20-80}$ estimates of the \cite{Afshari2016} duration GMM were evaluated as candidates for $D_{gm}$, but the $D_{a0.05-0.85}$ of AS96 resulted to a better fit of $PSA_{NGA}$.
The results of this comparison can be found in the Electronic supplement, Section S2. 


The AS96 functional form for the mean estimate or the $D_{0.05-0.75}$ duration is:

\begin{equation}
    \ln{D_{5-75}} \left\{ \begin{array}{lllll}
        \ln \left( \frac{1}{f_c} + c_1 (R_{rup} - R_c) + c_2 S \right) & for& R_{rup} \ge R_c \\
        \ln \left( \frac{1}{f_c} + c_2 S                       \right) & for& R_{rup} <  R_c \\
    \end{array} 
    \right. 
\end{equation}

\noindent where $f_c$ is the corner frequency of the earthquake:

\begin{equation} \label{eq:f_c}
    f_c = 4.9~10^6 \left( \frac{\Delta \sigma}{10^{1.5 M + 16.05}} \right)
\end{equation}

\noindent $\beta$ is the shear-wave velocity at the source, and $\Delta \sigma$ is the stress drop. 
$1/f_c$ is the source duration, $c_1 (R_{rup} - R_c)$ captures the distance dependence, and $c_2 S$ captures the site dependence. 
The scaling of AS96 has a physical basis because the distance and site dependence terms are additive, instead of multiplicative, to the source duration.
The rational for an additive distance dependence is that small and large magnitude earthquakes are expected to have a similar increase of duration with increasing distance due to the scattering of the seismic waves.
Similarly, the duration increase due to the site effects is also expected to be independent of the earthquake size. 
In AS96, other interval of significant duration can be calculated with Equation \eqref{eq:sig_dur_other}.

\begin{equation} \label{eq:sig_dur_other}
    \ln\left(\frac{D_{0.05-I}}{D_{5-75}}\right) = a_1 + a_2  \ln\left(\frac{I-0.05}{1-I}\right) + a_3  \ln\left(\frac{I-0.05}{1-I}\right)^2
\end{equation}

\subsubsection{Correction for non-stationarity} \label{sec:dur_non_stat}

One of RVT's main assumptions that is violated when applied in ground motions is that the signal is stationary. 
Especially when predicting $PSA$ for large $T_0$, an SDOF oscillator will not abruptly stop at the end of the ground motion, instead it will have a transient decaying response, which if not considered, would lead to an overestimation of $x_{rms}$.
To solve this problem, \cite{Boore1984} (JB84) proposed to include the oscillator duration ($D_o$) in $D_{rms}$ as shown in Equation \eqref{eq:dur_rms}; $D_o$ is not included in the calculation of the $PF$ because the response of the oscillator follows a steady decay after the end of the excitation. 
\cite{Liu1999} (LP99) improved the estimate of $D_o$ by considering the spectral shape of the input time history in the $D_o$ scaling.
\cite{Boore2012} (BT12), and \cite{Boore2015} (BT15) proposed a relationship for $D_{rms}/D_{gm}$; they used a more flexible functional form compared to the previous studies and considered the magnitude and distance scaling of $D_{rms}/D_{gm}$.

\begin{equation} \label{eq:dur_rms}
    D_{rms} = D_{gm} + D_o
\end{equation}

The BT15 oscillator duration model was selected for the subsequent analyses, as in preliminary evaluations, the RVT $PSA$ estimates with BT15 provided a better fit to the recorded $PSA$ than the alternative models.
Although BT12 performed equally well in estimating the $PSA$ of medium-to-large earthquakes, it was not selected because its is not applicable to magnitudes less than $4$.

\subsubsection{Extrapolation of EAS} \label{sec:eas_extrap}

To ensure that entire frequency content of the ground-motion is captured in the RVT calculations, both the ergodic and non-ergodic $EAS$ spectra are extrapolated at low and high frequencies. 
At low frequencies, $EAS$ is extrapolated to $0.01 Hz$ with an omega-square model \citep{Brune1970}:

\begin{equation}
\begin{aligned}
    &\Omega(f) = \frac{f^2}{ 1 + f^2/f_c^2 } \\
    &EAS(f < f_{min}) = A_{f_{min}} \Omega(f)
\end{aligned}
\end{equation}

\noindent where $f_c$ is the corner frequency (Equation \eqref{eq:f_c}), and $A_{f_{min}}$ is the amplitude of the omega-squared model at the minimum frequency of the $EAS$ ($f_{min}$).
The stress drop for the calculation of $f_c$ for the omega-squared model is estimated with the \cite{Atkinson2011} empirical relationship.
$A_{f_{min}}$ is estimated based on the $EAS$ amplitudes of $1.00 f_{min}$ to $1.05 f_{min}$ frequency bin:

\begin{equation}
    A_{f_{min}} = mean\left( \frac{EAS(f)}{\Omega(f)} \right) \quad for ~ f \in [1.0 f_{min}, 1.05 f_{min}]
\end{equation}

At high frequencies, $EAS$ is extrapolated to $100 Hz$ with a kappa model \citep{Anderson1984}:
\begin{equation}
\begin{aligned}
    &D(f) = exp(- \pi \kappa f)\\
    &EAS(f > f_{max}) = A_{f_{max}} D(f)
\end{aligned}
\end{equation}

\noindent $\kappa$ defines the rate of decay of the high frequencies, and $ A_{f_{max}}$ is the amplitude of the kappa model at the largest $EAS$ frequency, $f_{max}$.
$\kappa$ can be estimated with the \cite{Ktenidou2014} $\kappa - V_{S30}$ empirical relationship:

\begin{equation} \label{eq:kappa_vs30}
    \ln( \kappa ) = -0.4 \ln\left( \frac{V_{S30}}{760} \right) - 3.5
\end{equation}

$A_{f_{max}}$ is estimated based on the $EAS$ amplitudes in the $0.95 f_{max}$ to $1.00 f_{max}$ frequency bin:

\begin{equation}
    A_{f_{max}} = mean\left( \frac{EAS(f)}{D(f)} \right) \quad for ~ f \in [0.95 f_{max}, 1.00 f_{max}]
\end{equation}

\begin{figure}
    \centering
    \includegraphics[width=0.5\textwidth]{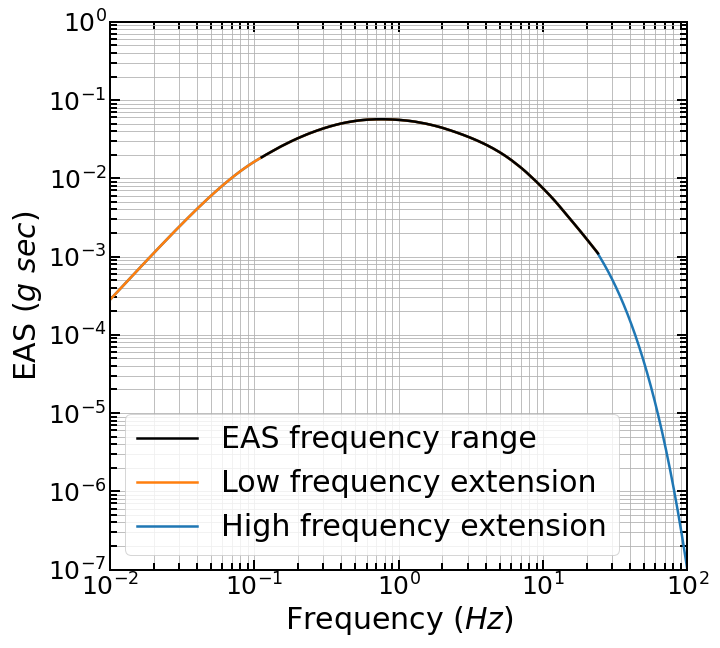}
    \caption{Extrapolation of $EAS$ to low and high frequencies. $EAS$ is estimated for $M=7$, $R_{rup}=30km$,and $V_{S30}=400m/sec$.}
    \label{fig:eas_extrap}
\end{figure}

As an example of the extrapolation procedure, the median estimate of the ergodic $EAS$ for a $M~7$ event, at a $R_{rup}$ distance of $30km$, and a $V_{S30}$ value of $400m/sec$ is extend to high and low frequencies using the omega-squared and kappa models in Figure \ref{fig:eas_extrap}, which shows that the amplitudes of the extended frequencies are in agreement with the $EAS$ over the usable frequency range.

\subsubsection{RVT summary and validation} \label{sec:RVT_summary}

In summary, all subsequent RVT calculations are performed with: the V75 $PF$, the median estimate of AS96 for $D_{a0.05-0.85}$ as $D_{gm}$, BT15 for $D_{rms}$, and the extrapolation procedure described in the previous subsection. 

\begin{figure}
    \centering
    \begin{subfigure}[t]{0.40\textwidth} 
        \caption{} 
        \includegraphics[width=.95\textwidth]{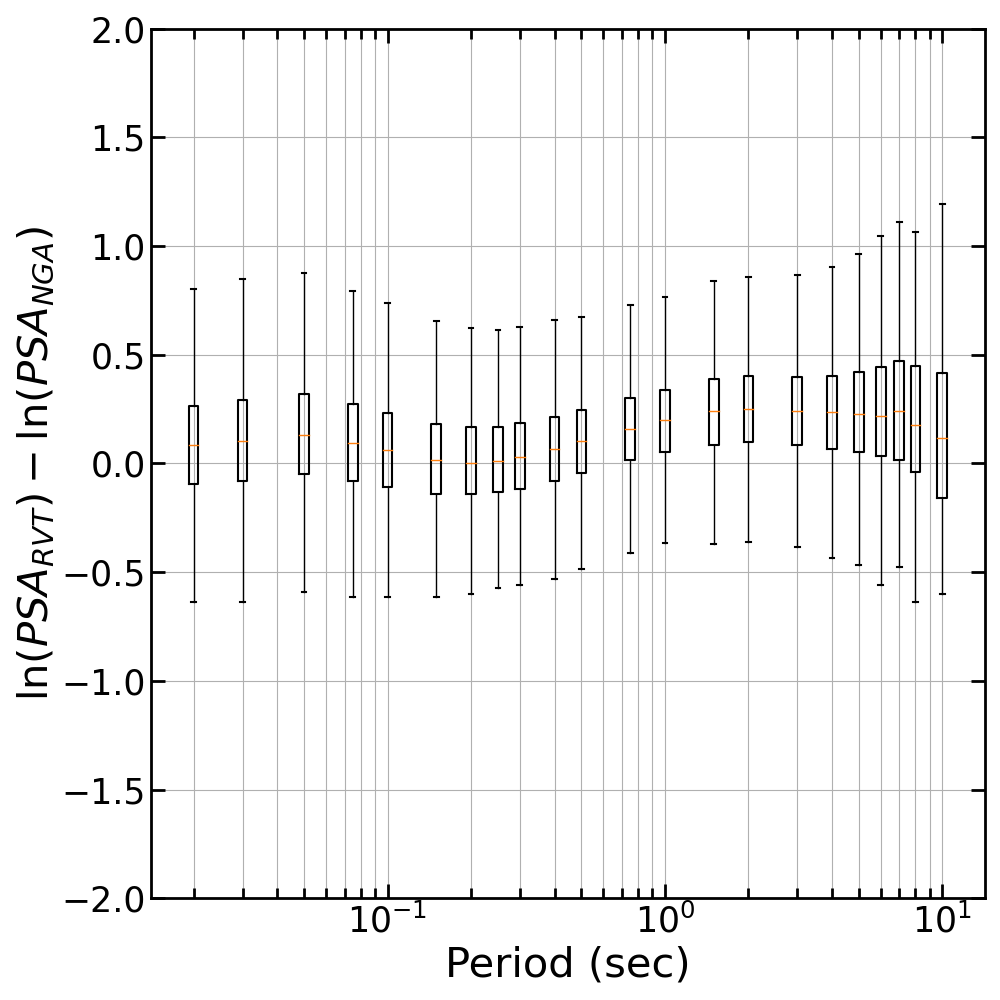}
    \end{subfigure}
    \begin{subfigure}[t]{0.40\textwidth} 
        \caption{} 
        \includegraphics[width=.95\textwidth]{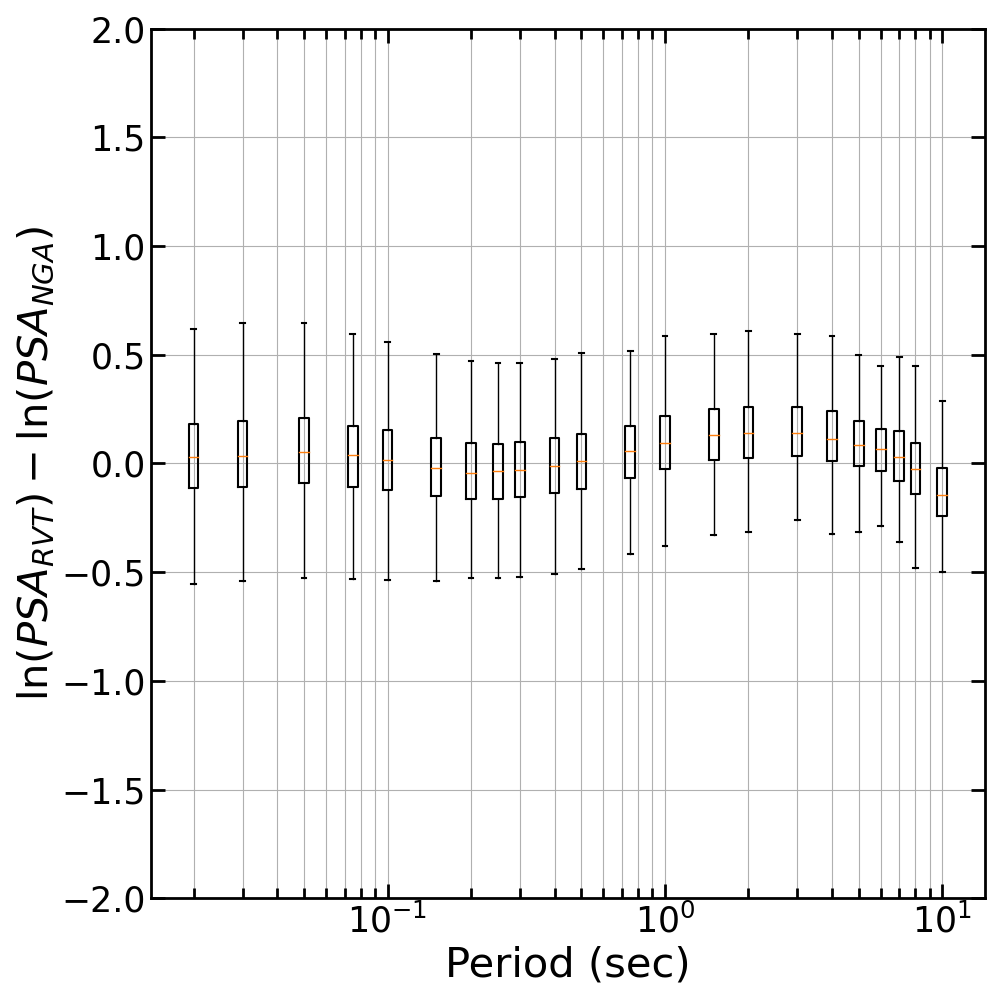}
    \end{subfigure}
    \caption{Residuals between the records' $PSA$ and $PSA$ calculated with RVT. (a) residuals of records of all $M$, (b) residuals of records of $M > 5$ }
    \label{fig:cmp_PF_BT15_dur_AS96}
\end{figure}

\begin{figure}
    \centering
    \begin{subfigure}[t]{0.40\textwidth} 
        \caption{} 
        \includegraphics[width=.95\textwidth]{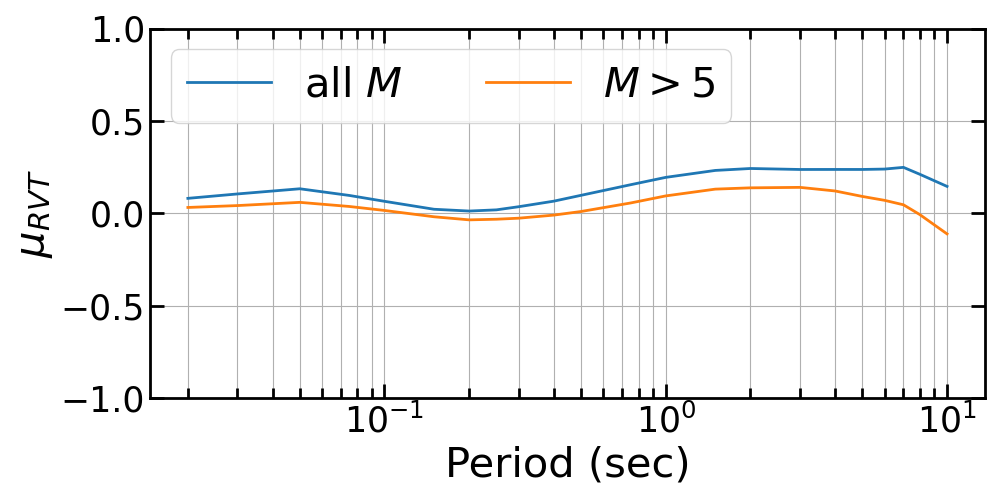}
    \end{subfigure}
    \begin{subfigure}[t]{0.40\textwidth} 
        \caption{} 
        \includegraphics[width=.95\textwidth]{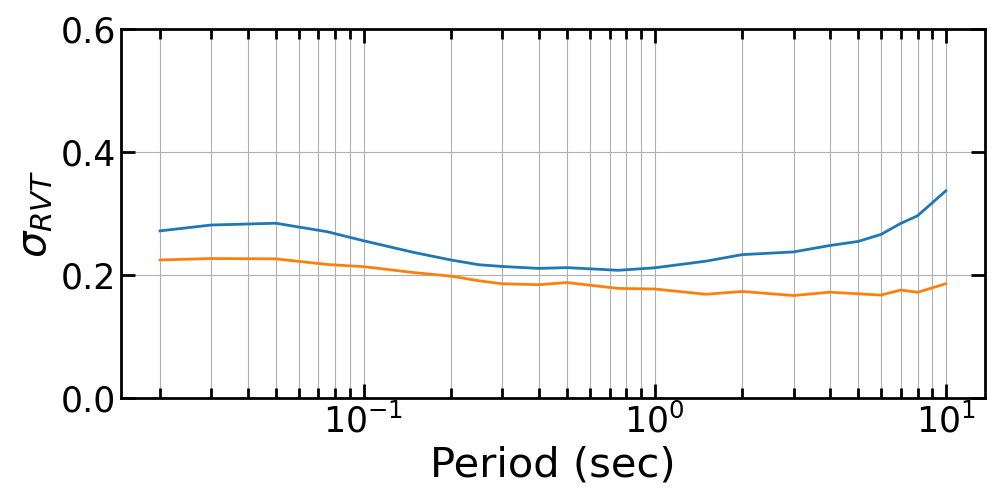}
    \end{subfigure}
    \caption{Mean and standard deviation of the residuals between the records' $PSA$ and $PSA$ calculated with RVT.}
    \label{fig:cmp_PF_BT15_dur_AS96_mean_std}
\end{figure}

As a validation, Figure \ref{fig:cmp_PF_BT15_dur_AS96} shows the residuals between the natural-log of $PSA_{NGA}$ and the natural-log of $PSA_{RVT}$ with the recommended $RVT$ procedure. 
Overall, $PSA_{RVT}$ is in good agreement with $PSA_{NGA}$ for the entire period range ($T_0=0.01 - 10 sec$) with the fit improving for $M>5$.
Figure \ref{fig:cmp_PF_BT15_dur_AS96_mean_std} shows the mean and the standard deviation of the residuals versus $T_0$. 
The residuals have a positive bias at $T_0 = 1 - 4 sec$; however, this is not propagated in the non-ergodic $PSA$ GMM, as the GMM is developed using non-ergodic factors, which are defined in the next subsection (Section \ref{sec:psa_nerg_ratios}).
The standard deviation or the residuals is approximately $0.2$ natural-log units for the entire period range. 

\subsection{Non-ergodic $PSA$ factors} \label{sec:psa_nerg_ratios}

The non-ergodic effects of the proposed $PSA$ GMM are expressed in terms of a non-ergodic $PSA$ factor ($F_{nerg}$); that is, the difference of the logs the non-ergodic $PSA$ estimate for a scenario of interest over the ergodic $PSA$ estimate for the same scenario (Equation \eqref{eq:nerg_ratios})
The non-ergodic $PSA$ values are calculated with RVT and the \cite{Lavrentiadis2021} non-ergodic $EAS$ GMM (LAK21), and the ergodic $PSA$ values are calculated with RVT and the \cite{Bayless2019a} ergodic $EAS$ GMM (BA18). 
The scenarios of interest are defined by the magnitude ($M$), closest-rupture distance ($R_{rup}$), time-average shear-wave velocity at the top $30m$ ($V_{S30}$), etc., which are input parameters to both the ergodic and non-ergodic $EAS$ GMMs, but also the earthquake and site coordinates, $\vec{x}_{e}$ and $\vec{x}_{s}$, which define the source, path and site non-ergodic effects in LAK21.
In this formulation, $F_{nerg}$ captures the combined effect of all non-ergodic terms; there are no separate terms for the earthquake, path, and site non-ergodic effects.

\begin{equation} \label{eq:nerg_ratios}
\begin{aligned}
    F_{nerg}&(T_0,M,R_{rup},V_{S30},\vec{x}_{e},\vec{x}_{s},..) = \\
           =&  \ln\left( PSA_{RVT}\left[ IR(T_0) ~ EAS_{LAK21}(M,R_{rup},V_{S30},\vec{x}_{e},\vec{x}_{s},...) \right] \right) \\
            &- \ln\left( PSA_{RVT}\left[ IR(T_0) ~ EAS_{BA18}(M,R_{rup},V_{S30},...) \right] \right)
\end{aligned}
\end{equation}

The proposed non-ergodic $PSA$ GMM is developed by coupling the aforementioned non-ergodic  with an existing ergodic $PSA$ GMM: 

\begin{equation}
\begin{aligned}
    y_{nerg}&(M,R,V_{S30},\vec{x}_{e},\vec{x}_{s},...) = \\
           =& y_{erg}(M,R,V_{S30},...) + F_{nerg}(M,R,V_{S30},\vec{x}_{e},\vec{x}_{s},...)
\end{aligned}
\end{equation}

\noindent where $y_{nerg}$ is the natural log of the non-ergodic $PSA$ median estimate, and $y_{erg}$ is the natural log of the ergodic median estimate.
The benefit of this approach is that it separates the non-ergodic effects from the average ground-motion scaling. 
$F_{nerg}$ does not affect the average scaling of the non-ergodic $PSA$ GMM, as LAK21 is based on BA18, and thus, their average scaling  is canceled out.
Furthermore, the small bias of $RVT$ is also canceled out in this approach, as the same $RVT$ procedure is used to compute $PSA_{erg}$ and $PSA_{nerg}$
For the average scaling of the non-ergodic $PSA$ GMM, $y_{erg}$, we chose the \cite{Abrahamson2014} (ASK14) and \cite{ChiouYoungs2014} (CY14) ergodic $PSA$ GMMs.
Hereafter, the non-ergodic GMM that is based on ASK14 is called non-ergodic GMM\textsubscript{1}, and the non-ergodic GMM that is based on CY14 is called non-ergodic GMM\textsubscript{2}.
The main reasons ASK14 and CY14 are selected to develop the non-ergodic GMM are: i) they were developed with the same data-set as BA18, and ii) they include complex scaling terms, such as hanging-wall effects, which can be passed to the non-ergodic GMMs. 

\begin{figure}
    \centering
    \begin{subfigure}[t]{0.32\textwidth} 
        \caption{} 
        \includegraphics[width=.98\textwidth]{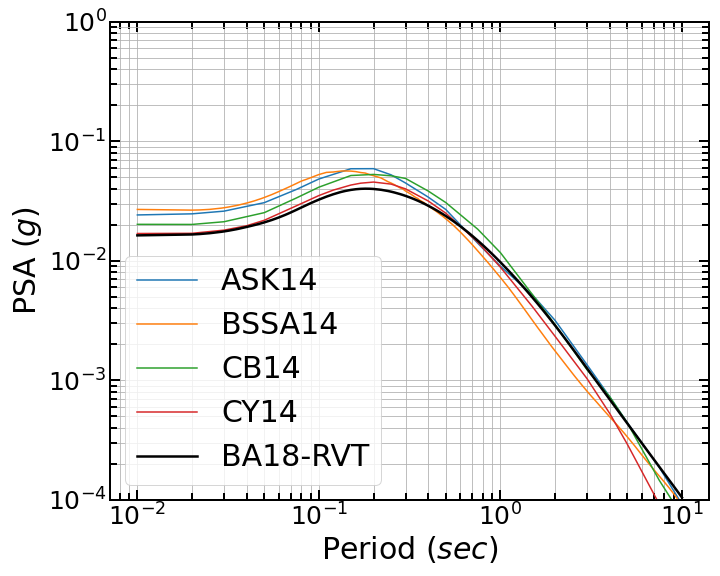}
    \end{subfigure}
    \begin{subfigure}[t]{0.32\textwidth}
        \caption{}  
        \includegraphics[width=.98\textwidth]{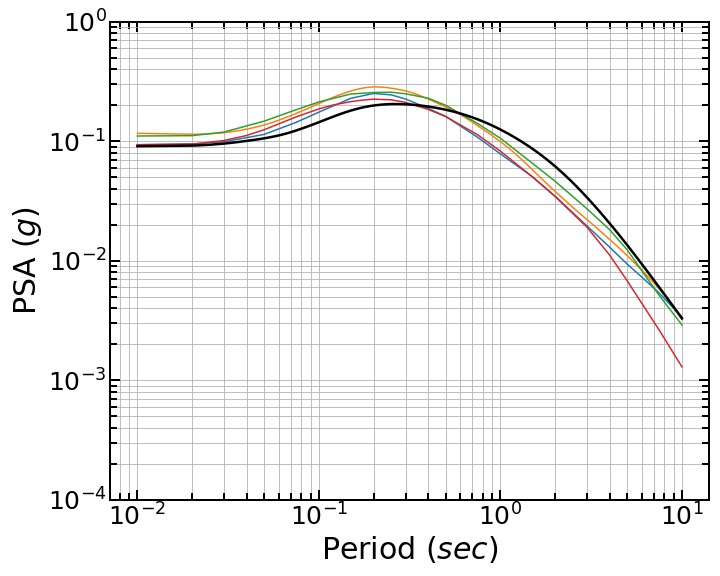}
    \end{subfigure}
    \begin{subfigure}[t]{0.32\textwidth}
        \caption{}  
        \includegraphics[width=.98\textwidth]{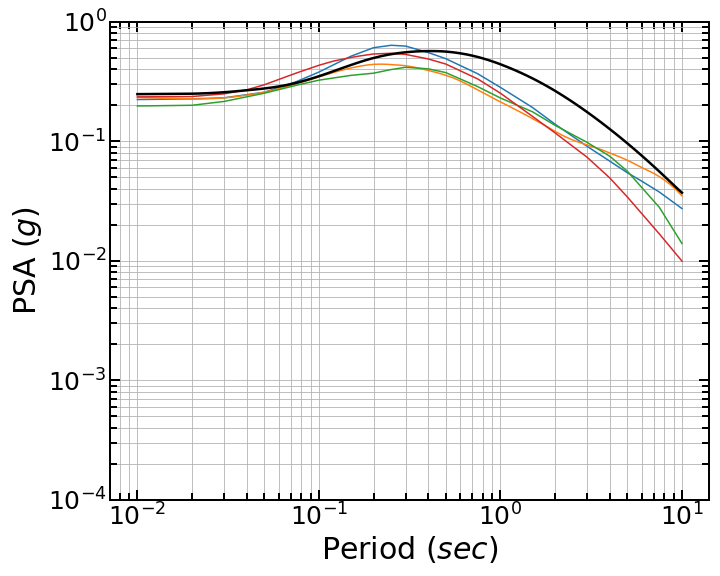}
    \end{subfigure}
    \caption{Comparison of $PSA$ spectra developed with the BA18 EAS GMM and RVT, shown with the black line, and $PSA$ spectra estimated using the NGAWest2 GMMs, shown with the colored lines. (a) $M~5.0$, (b) $M~6.5$, and (c) $M~8.0$ earthquake scenario with $R_{rup}=30~km$ and $V_{S30} = 400~m/sec$. }
    \label{fig:cmp_gmm_nga}
\end{figure}

The non-ergodic $PSA$ GMM was not developed directly with RVT and LAK21 because this approach  led to an overestimation the median $PSA$ at medium-to-large periods.
Figure \ref{fig:cmp_gmm_nga} compares the four NGAWest2 GMMs: ASK14, BSSA14, CB14, and CY14 \citep{Abrahamson2014, Boore2014, Campbell2014, ChiouYoungs2014} with the spectral acceleration response spectrum created with RVT and BA18.
The NGAWest2 GMMs are in good agreement with the $PSA$ from BA18 for the $M~5$ event, but the comparison worsens as the size of the earthquake increases. 
For periods $T_0=2-4sec$, for the $M~8$ earthquake, the $PSA$ from BA18 is a factor of two higher than the NGAWest2 GMMs, indicating that, in this period range, BA18 has a stronger magnitude scaling than the NGAWest2 GMMs. 
Since LAK21 is based on BA18, a non-ergodic $PSA$ GMM developed with RVT and LAK21 will also have a stronger magnitude scaling than the NGAWest2 GMMs.
Due to the effort involved in the development of the NGAWest2 GMMs, we judge that their magnitude scaling is more likely to be correct, which is why we used the non-ergodic factors approach to develop the non-ergodic GMM; however, future studies should further investigate the cause of the different magnitude scaling.  

The epistemic uncertainty of the non-ergodic $PSA$ GMM is captured by sampling the non-ergodic terms of LAK21 GMM multiple times and calculating the $F_{nerg}$ for each sample. 
As shown in the example in Section \ref{sec:examp_infreq}, it is important to consider the inter-frequency correlation of the non-ergodic terms, otherwise the epistemic uncertainty is underestimated. 

\subsection{Constant Shift and Aleatory Model}

The constant shift ($\delta c_0$), between-event residuals ($\delta B^0_{e}$), and within-event within-site residuals ($\delta WS^0_{e,s}$) are estimated by fitting a mixed-effects linear model to the total residuals of the non-ergodic models: 

\begin{equation} \label{eq:aleat_reg}
    \epsilon_{e,s} = \delta c_0 + \delta B^0_{e} + \delta WS^0_{e,s}
\end{equation}

\begin{figure}
    \centering
    \begin{subfigure}[t]{0.40\textwidth} 
        \caption{} 
        \includegraphics[width=.95\textwidth]{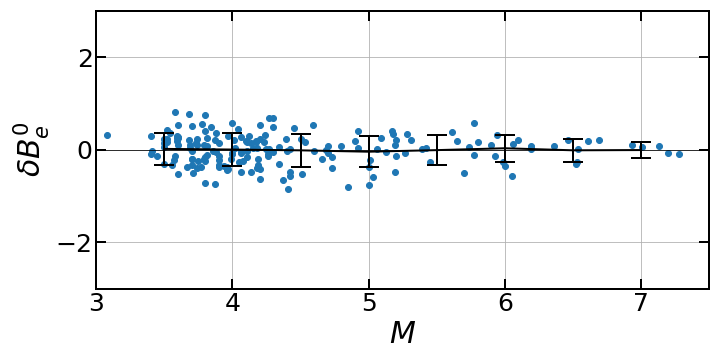}
    \end{subfigure}
    \begin{subfigure}[t]{0.40\textwidth} 
        \caption{} 
        \includegraphics[width=.95\textwidth]{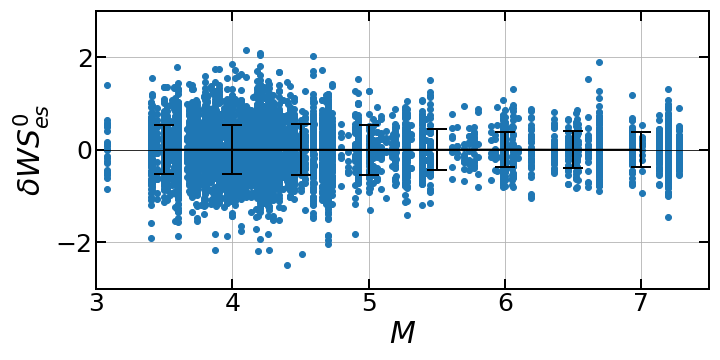}
    \end{subfigure}
    \begin{subfigure}[t]{0.40\textwidth} 
        \caption{} 
        \includegraphics[width=.95\textwidth]{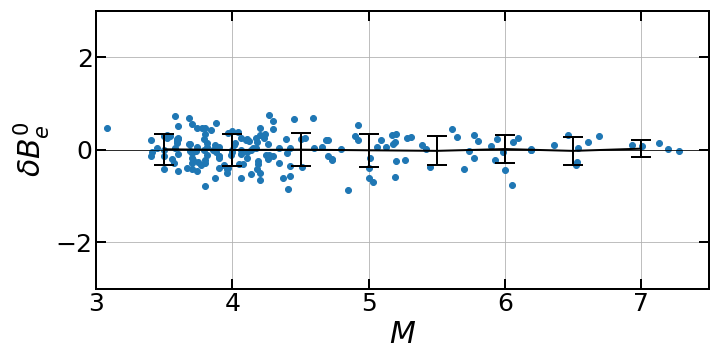}
    \end{subfigure}
    \begin{subfigure}[t]{0.40\textwidth} 
        \caption{} 
        \includegraphics[width=.95\textwidth]{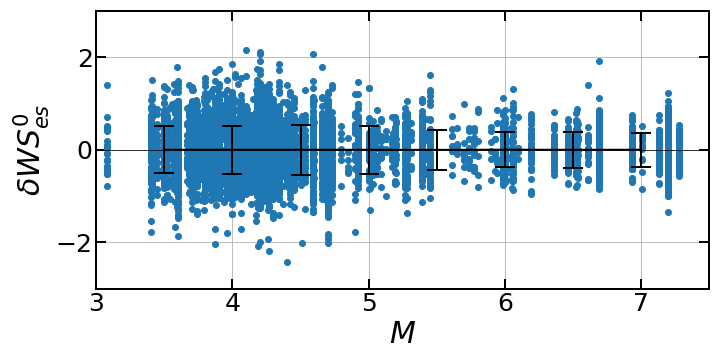}
    \end{subfigure}
    \caption{Between-event and within-event within-site residuals for $T_0=0.25sec$ versus magnitude. (a) $\delta B_e$ of non-ergodic GMM\textsubscript{1}, (b) $\delta WS_{e,s}$ of non-ergodic GMM\textsubscript{1}, (c) $\delta B_e$ of non-ergodic GMM\textsubscript{2}, and (d) $\delta WS_{e,s}$ of non-ergodic GMM\textsubscript{2}.}
    \label{fig:res_GMM1GMM2_vs_mag}
\end{figure}

\begin{figure}
    \centering
    \begin{subfigure}[t]{0.40\textwidth} 
        \caption{} 
        \includegraphics[width=.95\textwidth]{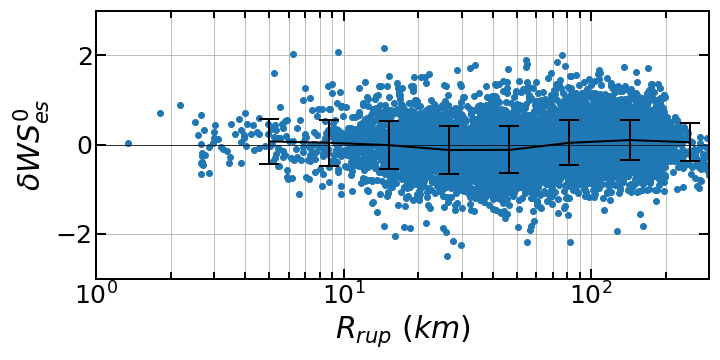}
    \end{subfigure}
    \begin{subfigure}[t]{0.40\textwidth} 
        \caption{} 
        \includegraphics[width=.95\textwidth]{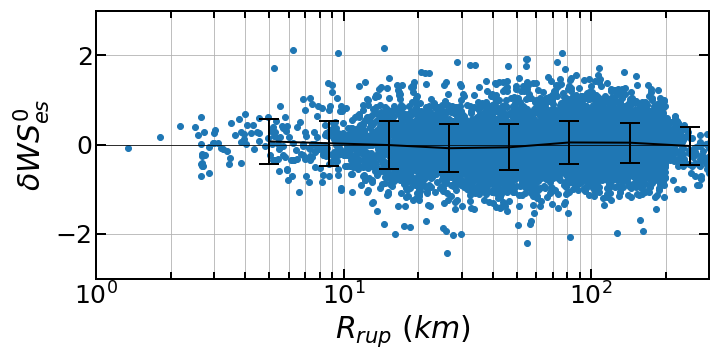}
    \end{subfigure}
    \caption{Within-event within-site residuals for $T_0=0.25sec$ versus $R_{rup}$. (a) $\delta WS_{e,s}$ of non-ergodic GMM\textsubscript{1}, (b) $\delta WS_{e,s}$ of non-ergodic GMM\textsubscript{2}.}
    \label{fig:res_GMM1GMM2_vs_rrup}
\end{figure}

\begin{figure}
    \centering
    \begin{subfigure}[t]{0.40\textwidth} 
        \caption{} 
        \includegraphics[width=.95\textwidth]{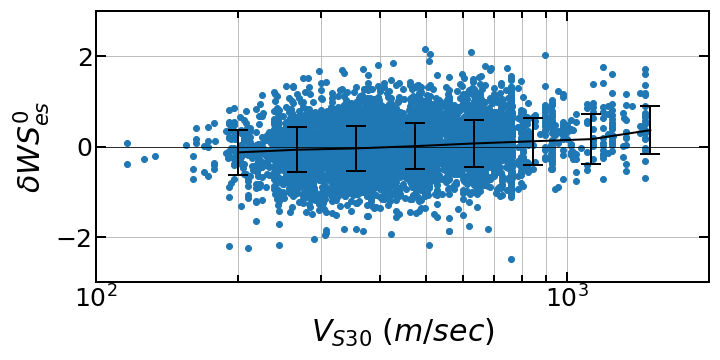}
    \end{subfigure}
    \begin{subfigure}[t]{0.40\textwidth} 
        \caption{} 
        \includegraphics[width=.95\textwidth]{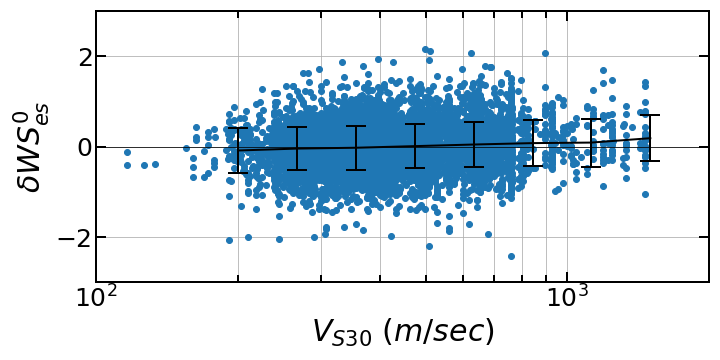}
    \end{subfigure}
    \caption{Within-event within-site residuals for $T_0=0.25sec$ versus $V_{S30}$. (a) $\delta WS_{e,s}$ of non-ergodic GMM\textsubscript{1}, (b) $\delta WS_{e,s}$ of non-ergodic GMM\textsubscript{2}.}
    \label{fig:res_GMM1GMM2_vs_vs30}
\end{figure}

The magnitude dependence of $\delta B^0_{e}$ and  $\delta WS^0_{e,s}$ of the two non-ergodic $PSA$ GMMs for $T_0=0.25sec$ is evaluated in Figure \ref{fig:res_GMM1GMM2_vs_mag}.
The mean of $\delta B^0_{e}$ and $\delta WS^0_{e,s}$ shows no trend with $M$, but their empirical standard deviation decreases with $M$. 
Similarly, the $R_{rup}$ and $V_{S30}$ dependence of the $\delta WS^0_{e,s}$ for $T_0=0.25sec$ is evaluated in Figures \ref{fig:res_GMM1GMM2_vs_rrup} and \ref{fig:res_GMM1GMM2_vs_vs30} where no significant trends are found in either the mean or the standard deviation.


\begin{figure}
    \centering
    \begin{subfigure}[t]{0.40\textwidth} 
        \caption{} 
        \includegraphics[width=.95\textwidth]{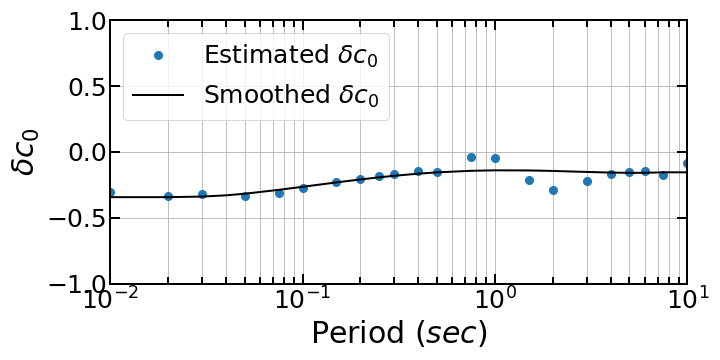}
    \end{subfigure}
    \begin{subfigure}[t]{0.40\textwidth} 
        \caption{} 
        \includegraphics[width=.95\textwidth]{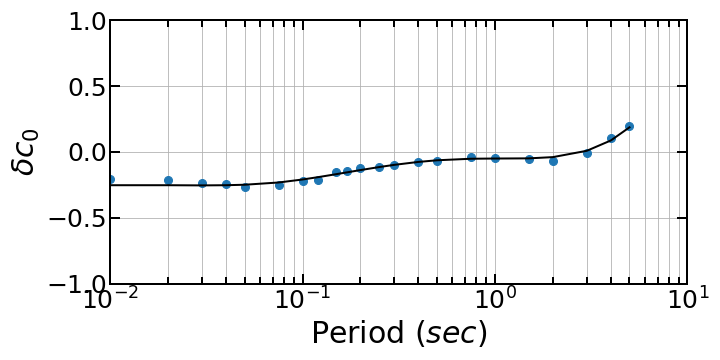}
    \end{subfigure}
    \caption{Estimated and smoothed $\delta c_0$ versus $T_0$. (a) non-ergodic GMM\textsubscript{1}, (b) non-ergodic GMM\textsubscript{2}}
    \label{fig:c0_model}
\end{figure}

Figure \ref{fig:c0_model} shows the estimated and smoothed $\delta c_0$ of the two non-ergodic $PSA$ GMMs.
Non-ergodic GMM\textsubscript{2}, which is based on CY14, is only estimated up to $T_o = 5~sec$ because, at larger periods, $\delta c_0$ deviated significantly from zero. 

\begin{figure}
    \centering
    \begin{subfigure}[t]{0.40\textwidth} 
        \caption{} 
        \includegraphics[width=.95\textwidth]{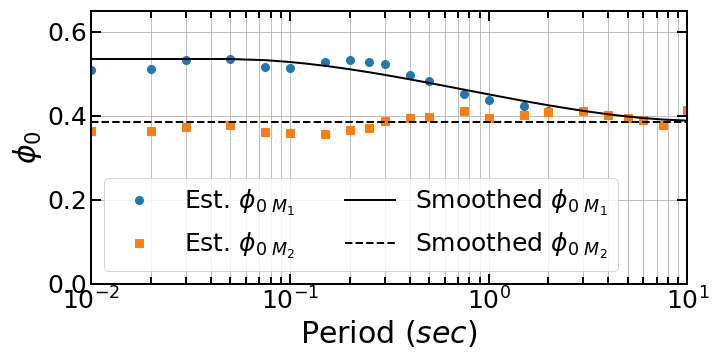}
    \end{subfigure}
    \begin{subfigure}[t]{0.40\textwidth} 
        \caption{} 
        \includegraphics[width=.95\textwidth]{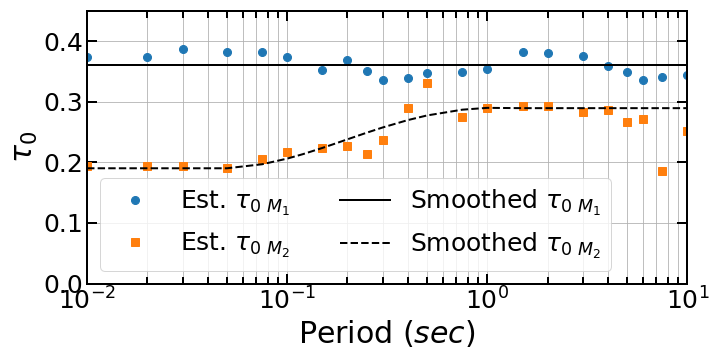}
    \end{subfigure}
    \begin{subfigure}[t]{0.40\textwidth} 
        \caption{} 
        \includegraphics[width=.95\textwidth]{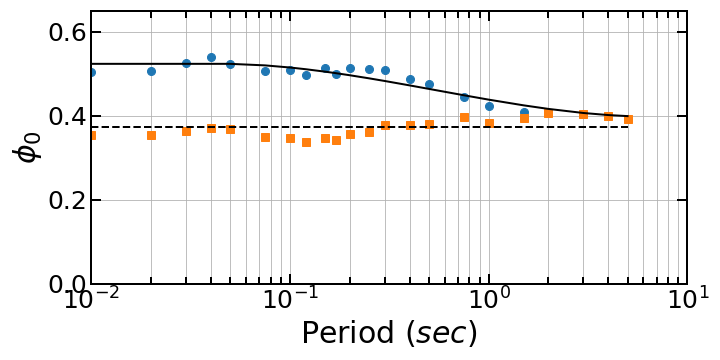}
    \end{subfigure}
    \begin{subfigure}[t]{0.40\textwidth} 
        \caption{} 
        \includegraphics[width=.95\textwidth]{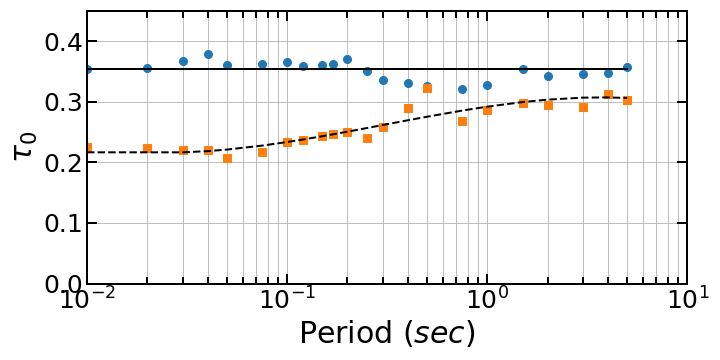}
    \end{subfigure}
    \caption{Period dependence of aleatory model parameters. 
    (a) period dependence of $\phi_{0M_1}$, $\phi_{0M_1}$ for non-ergodic GMM\textsubscript{1}
    (b) period dependence of $\tau_{0M_1}$, $\phi_{0M_1}$ for non-ergodic GMM\textsubscript{1}
    (c) period dependence of $\phi_{0M_1}$, $\phi_{0M_1}$ for non-ergodic GMM\textsubscript{2}
    (d) period dependence of $\tau_{0M_1}$, $\phi_{0M_1}$ for non-ergodic GMM\textsubscript{2}}
    \label{fig:phi0_tau0_model}
\end{figure}

Based on the empirical standard deviation of the non-ergodic residuals (Figure \ref{fig:res_GMM1GMM2_vs_mag}), both $\phi_0$ and $\tau_0$ are modeled as magnitude dependent (Equation \eqref{eq:phi_model} and \eqref{eq:tau_model}).
Figure \ref{fig:phi0_tau0_model} shows the period dependence of $\phi_0$ and $\tau_0$ for small and large magnitudes. 
The magnitude dependence of $\phi_0$ and $\tau_0$ is more significant at small periods.
The increase of the within-event aleatory variability at the small periods of small magnitudes may be caused by the radiation pattern which make the amplitude of the ground motion sensitive to the azimuthal angle.
For large magnitudes, which can be thought as many small events, the radiation patterns have less impact on the ground-motion variability, because the individual radiation patterns destructively interfere with each other due to the different azimuthal angles. 
Similarly, the larger between-event aleatory variability at the small periods of small magnitudes is believed to be caused by differences in stress drop which shifts the ground motions at frequencies above the corner frequency of the earthquake. 
Due to the larger rupture dimensions of the large events, any variability in the stress drop along the rupture averages out resulting in reduced between-event variability. 

The total standard deviation of the two non-ergodic GMMs are $30$ to $35\%$ smaller than the total standard deviation of the ergodic GMMs. 

\begin{equation} \label{eq:phi_model}
    \phi_0 = \left\{ \begin{array}{cll}
                \phi_{0M_1}                                             & for & M < 5  \\
                \phi_{0M_1} + (\phi_{0M_2}-\phi_{0M_2}) (M-5)/(6.5-5)   & for & 5 < M < 6.5  \\
                \phi_{0M_2}                                             & for & M > 6.5  \\
    \end{array} 
    \right. 
\end{equation}

\begin{equation} \label{eq:tau_model}
    \tau_0 = \left\{ \begin{array}{cll}
                \tau_{0M_1}                                             & for & M < 5  \\
                \tau_{0M_1} + (\tau_{0M_2}-\tau_{0M_2}) (M-5)/(6.5-5)   & for & 5 < M < 6.5  \\
                \tau_{0M_2}                                             & for & M > 6.5  \\
    \end{array} 
    \right. 
\end{equation}

\begin{figure}
    \centering
    \begin{subfigure}[t]{0.40\textwidth} 
        \caption{} 
        \includegraphics[width=.95\textwidth]{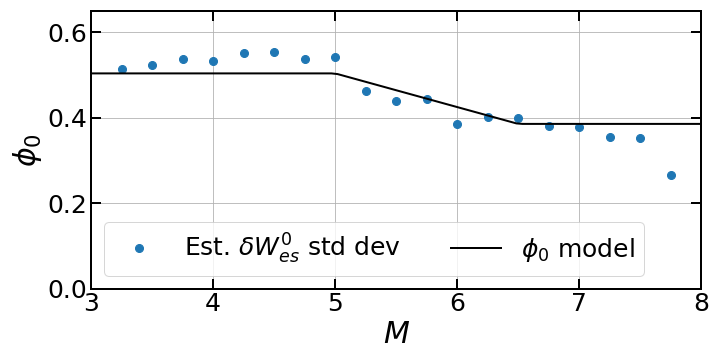}
    \end{subfigure}
    \begin{subfigure}[t]{0.40\textwidth} 
        \caption{} 
        \includegraphics[width=.95\textwidth]{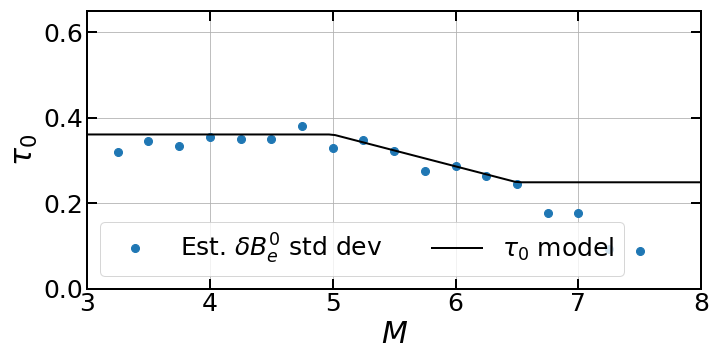}
    \end{subfigure}
    \begin{subfigure}[t]{0.40\textwidth} 
        \caption{} 
        \includegraphics[width=.95\textwidth]{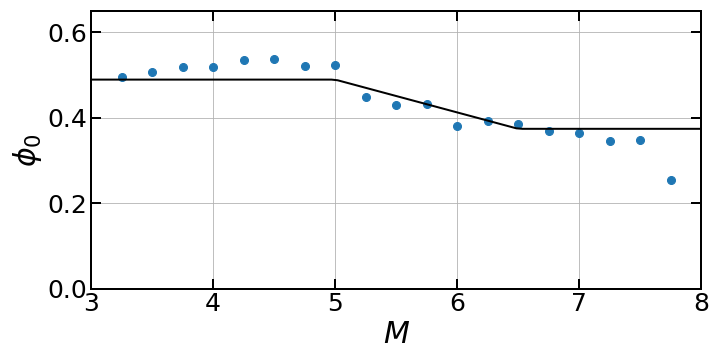}
    \end{subfigure}
    \begin{subfigure}[t]{0.40\textwidth} 
        \caption{} 
        \includegraphics[width=.95\textwidth]{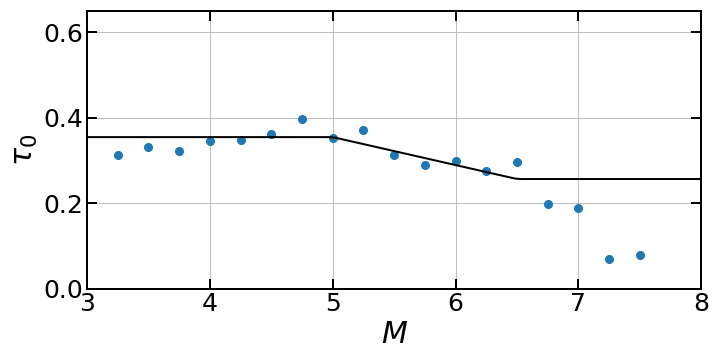}
    \end{subfigure}
    \caption{Magnitude dependence of $\phi_0$ and $\tau_0$ for $T_0=0.25sec$. 
    Circular markers denote the standard deviations of the binned residuals, and solid lines correspond to the standard deviation models. (a) $\phi_0$ for non-ergodic GMM\textsubscript{1}, (b) $\tau_0$ for non-ergodic GMM\textsubscript{1}, (a) $\phi_0$ for non-ergodic GMM\textsubscript{2}, and (b) $\tau_0$ for non-ergodic GMM\textsubscript{2}}
    \label{fig:alet_vs_mag}
\end{figure}

Figure \ref{fig:alet_vs_mag} compares the proposed models for $\phi_0$ and $\tau_0$ with the standard deviations of the binned residuals for $T_0=0.25 sec$. 
Overall, the aleatory models are in good agreement with the empirical standard deviations. 
The discrepancy at large magnitudes is considered acceptable, as the number of large magnitude events is small to reliably estimate the empirically standard deviation.

\begin{figure}
    \centering
    \begin{subfigure}[t]{0.40\textwidth} 
        \caption{} 
        \includegraphics[width=.95\textwidth]{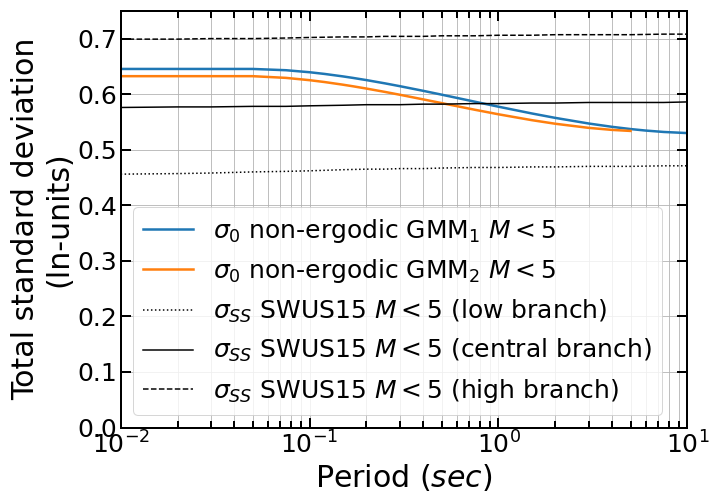}
    \end{subfigure}
    \begin{subfigure}[t]{0.40\textwidth} 
        \caption{} 
        \includegraphics[width=.95\textwidth]{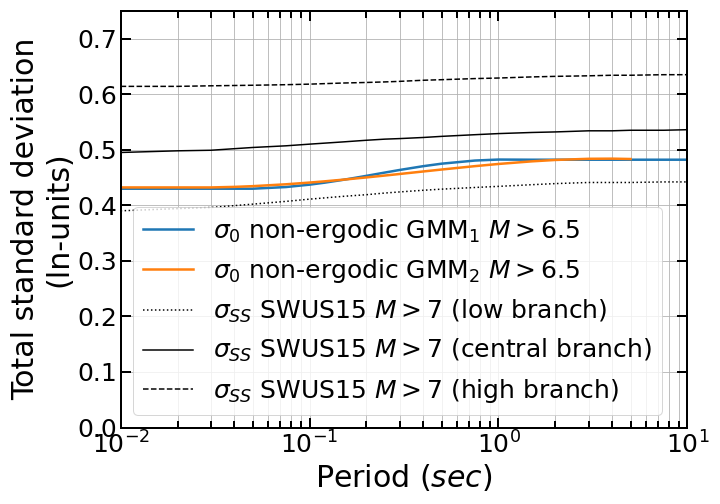}
    \end{subfigure}
    \caption{Comparison of total standard deviation of non-ergodic GMM\textsubscript{1} and GMM\textsubscript{2} with total standard deviation of SWUS15 partially non-ergodic GMM. 
    (a) small-to-moderate magnitude comparison, and (b) large magnitude comparison}
    \label{fig:cmp_sig}
\end{figure}

As a comparison with previous non-ergodic models, Figure \ref{fig:cmp_sig} shows the total standard deviation of the two non-ergodic GMMs and the total standard deviation of the SWUS15 partially non-ergodic GMM \citep{SWUS2015}. 
The standard deviations of non-ergodic GMM\textsubscript{1} and GMM\textsubscript{2} are within the low and high branches of SWUS15 for entire period range for both small-to-moderate and large events. 
For small-to-moderate magnitude events and $T_0 < 1sec$, the total standard deviations of GMM\textsubscript{1} and GMM\textsubscript{2} are larger than the median branch of SWUS15.
One possible reason for this is that $\sigma_{SS}$ of SWUS15 was estimated with magnitudes greater than $4$, whereas $\sigma_0$ of GMM\textsubscript{1} and GMM\textsubscript{2} were estimated with magnitudes greater than $3$ which exhibit larger variability at small periods. 
At large events, the total standard deviations of GMM\textsubscript{1} and GMM\textsubscript{2} are between the central and lower branch of SWUS15.
The GMM\textsubscript{1} and GMM\textsubscript{2} $\sigma_0$ values are expected to be less than SWUS15 $\sigma_{SS}$ central branch because in addition to the systematic site effects, GMM\textsubscript{1} and GMM\textsubscript{2} capture the systematic source and path effects; however, the fact that the $\sigma_0$ GMM\textsubscript{1} and GMM\textsubscript{2} are larger than the lower branch of SWUS15 means that the majority of the systematic effects captured by GMM\textsubscript{1} and GMM\textsubscript{2} are related to the site effects. 

\section{Applications} \label{sec:applications}

\subsection{Effect of $EAS$ inter-frequency correlation in $F_{nerg~PSA}$} \label{sec:examp_infreq}

In most GMMs, the ground-motion amplitude (i.e. $PSA$ or $EAS$) at every frequency is estimated independently; however, an actual ground-motion recording has peaks and troughs. That is the amplitudes of neighbouring frequencies are correlated. 
For instance, if amplitude of some frequency is above the average, it is likely that amplitudes of the nearby frequencies will also be above the average. 
This inter-frequency correlation is important in RVT, as the response of an SDOF oscillator does not only depend on the ground-motion amplitude at $T_0$ but also at the frequency content around $T_0$. 
\cite{Bayless2018} showed that the $PSA$ variability is underestimated if the inter-frequency correlation of $FAS$ is not considered. 

\begin{figure}
    \centering
    \begin{subfigure}[t]{0.40\textwidth} 
        \caption{} \label{fig:eas_var_ifindp}
        \includegraphics[width=.95\textwidth]{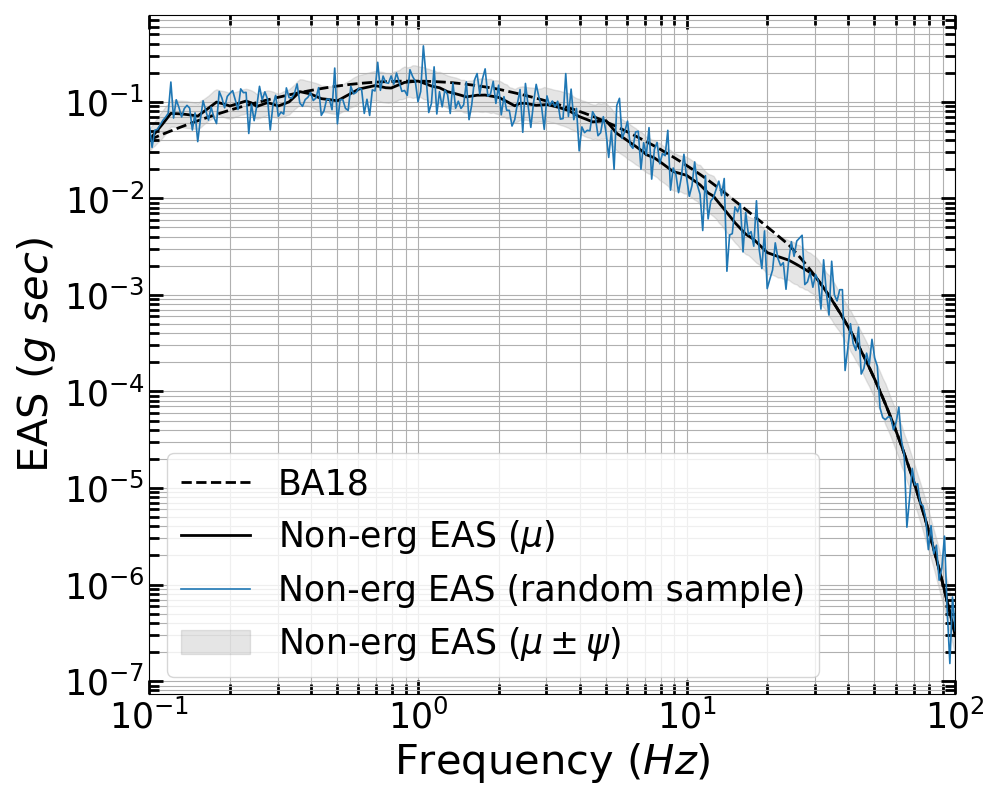}
    \end{subfigure}
    \begin{subfigure}[t]{0.40\textwidth}
        \caption{} \label{fig:eas_var_ifcorr}
        \includegraphics[width=.95\textwidth]{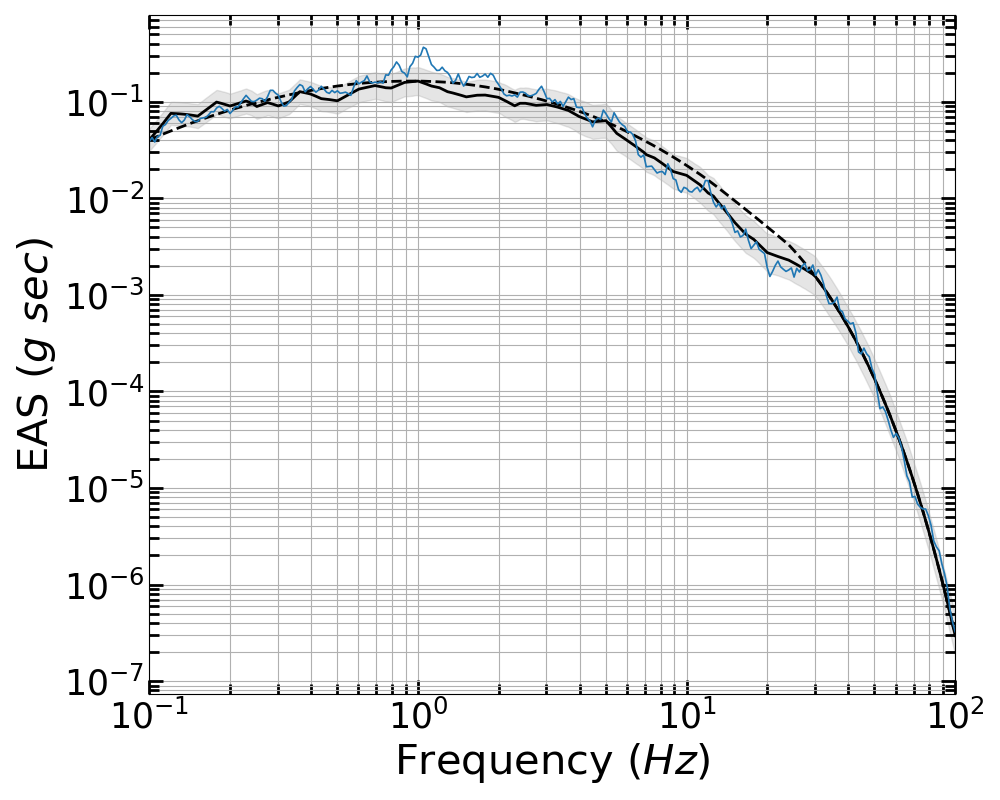}
    \end{subfigure}
    \caption{Effective amplitude spectra for a $M~7$ earthquake in Hayward fault, $8~km$ away from a site located in Berkeley CA.
    (a) without inter-frequency correlation, and 
    (b) with inter-frequency correlation. }
    \label{fig:eas_var}
\end{figure}

\begin{figure}
    \centering
    \begin{subfigure}[t]{0.40\textwidth} 
        \caption{} 
        \includegraphics[width=.95\textwidth]{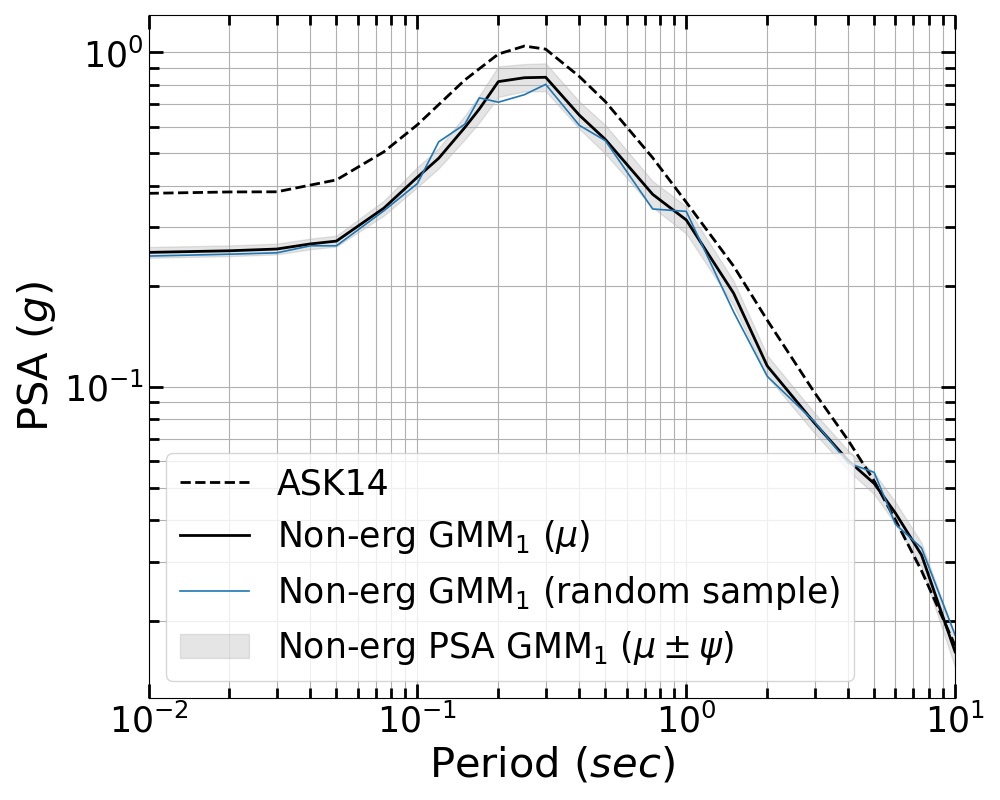}
    \end{subfigure}
    \begin{subfigure}[t]{0.40\textwidth}
        \caption{}  
        \includegraphics[width=.95\textwidth]{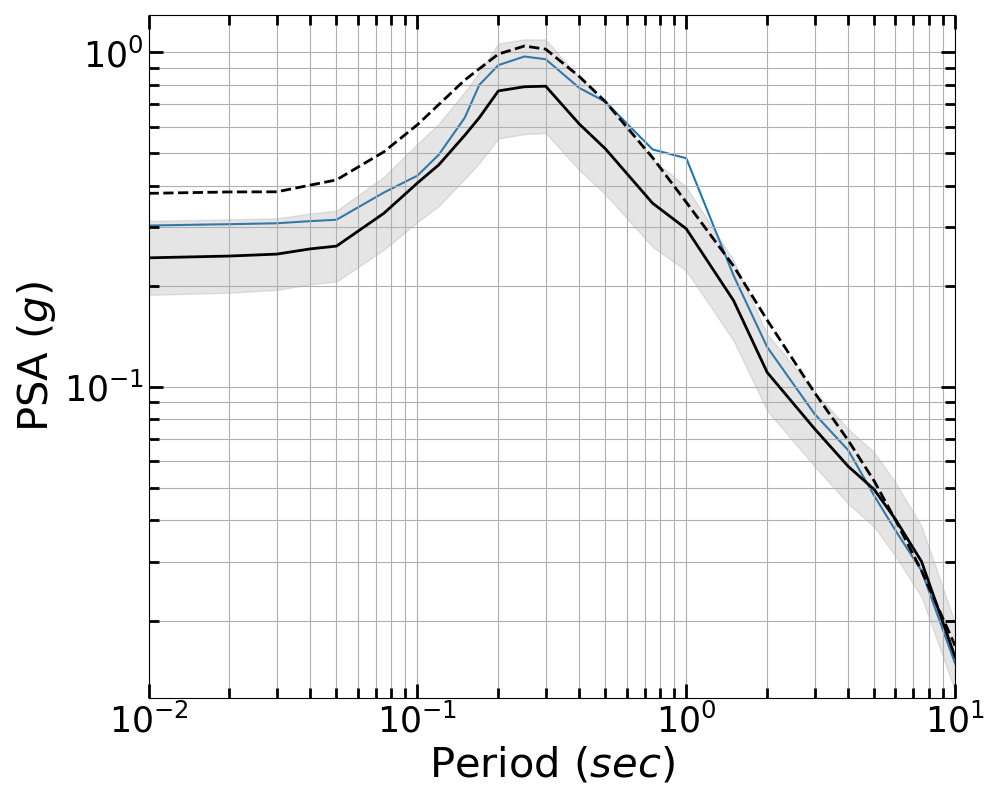}
    \end{subfigure}
    \caption{$PSA$ spectra for a $M~7$ earthquake in Hayward fault, $8~km$ away from a site located in Berkeley CA. 
    (a) without inter-frequency correlation, and 
    (b) with inter-frequency correlation. }
    \label{fig:psa_var}
\end{figure}

To illustrate the effect of the inter-frequency correlation in the calculation of $F_{nerg~PSA}$, we applied the proposed non-ergodic GMM with and without the inter-frequency correlation in $EAS$.
In both cases, the scenario of interest is a $M 7$ earthquake in Hayward Fault $8 km$ away from a site in Berkeley, CA.
The ergodic and non-ergodic $EAS$ of the two approaches are shown in Figure \ref{fig:eas_var}, and the corresponding non-ergodic $PSA$ spectra are shown in Figure \ref{fig:psa_var}.
The non-ergodic $EAS$ in Figure \ref{fig:eas_var_ifindp} are developed without inter-frequency correlation, whereas the non-ergodic $EAS$ in figure \ref{fig:eas_var_ifcorr} are developed using the inter-frequency correlation model in \cite{Lavrentiadis2021}.

In $EAS$ space, both approaches resulted in the same median and epistemic uncertainty range, but in $PSA$ space, only the median is the same.
 The epistemic uncertainty of $PSA$ is larger when the $EAS$ inter-frequency correlation is considered, because if $EAS$ is at an extreme at $T_0$ it will generally stay at the extreme over the neighbouring frequencies; thus, all the frequencies which influence the response of the oscillator will constructively interfere leading to a range of $PSA$ amplitudes that is wider. 
In contrast, if the $EAS$ amplitudes are uncorrelated, they will have negating effect on the response of the oscillator, resulting in a narrower range of $PSA$.
This shows the importance of considering the $EAS$ inter-frequency correlation in the non-ergodic $PSA$ calculations, as otherwise, the epistemic uncertainty of the $PSA$ is underestimated. 

\subsection{Magnitude dependence $F_{nerg~PSA}$}

\begin{figure}
    \centering
    \begin{subfigure}[t]{0.40\textwidth} 
        \caption{} 
        \includegraphics[width=.95\textwidth]{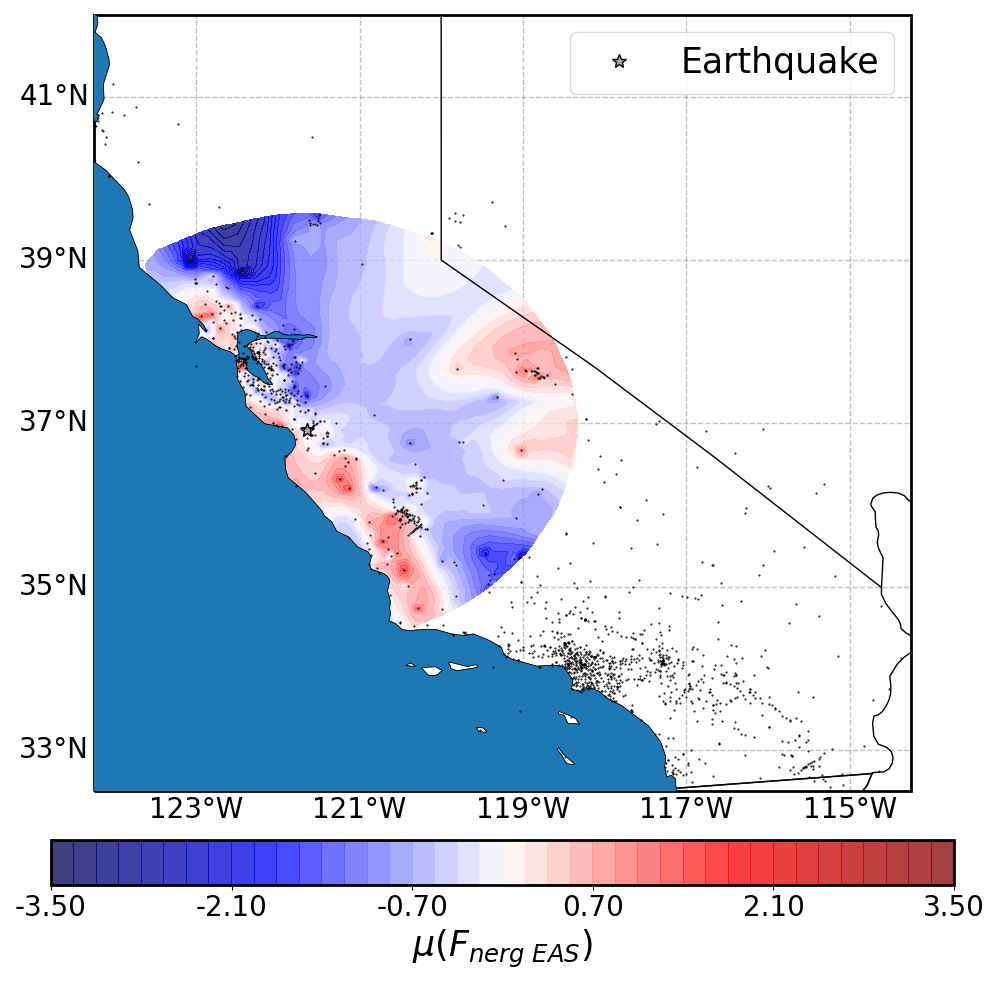}
    \end{subfigure}
    \begin{subfigure}[t]{0.40\textwidth}
        \caption{}  
        \includegraphics[width=.95\textwidth]{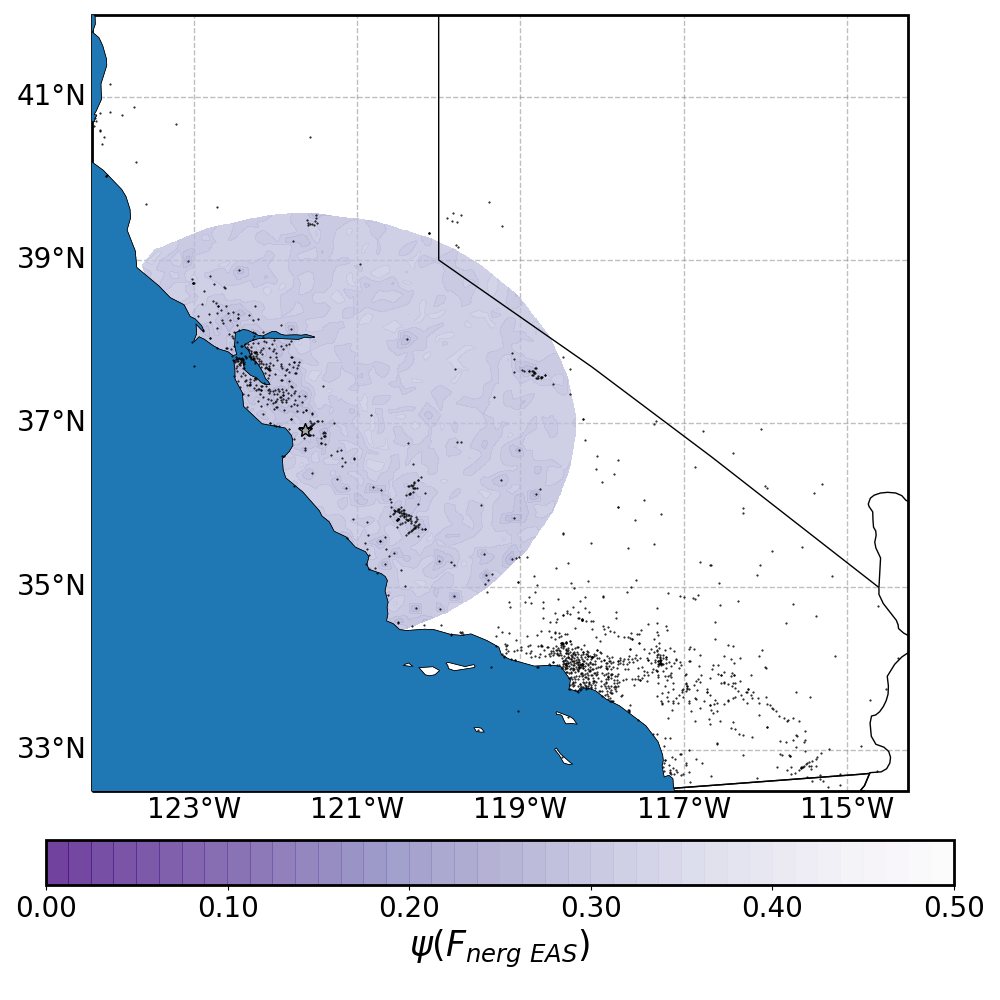}
    \end{subfigure}
    \begin{subfigure}[t]{0.40\textwidth} 
        \caption{} 
        \includegraphics[width=.95\textwidth]{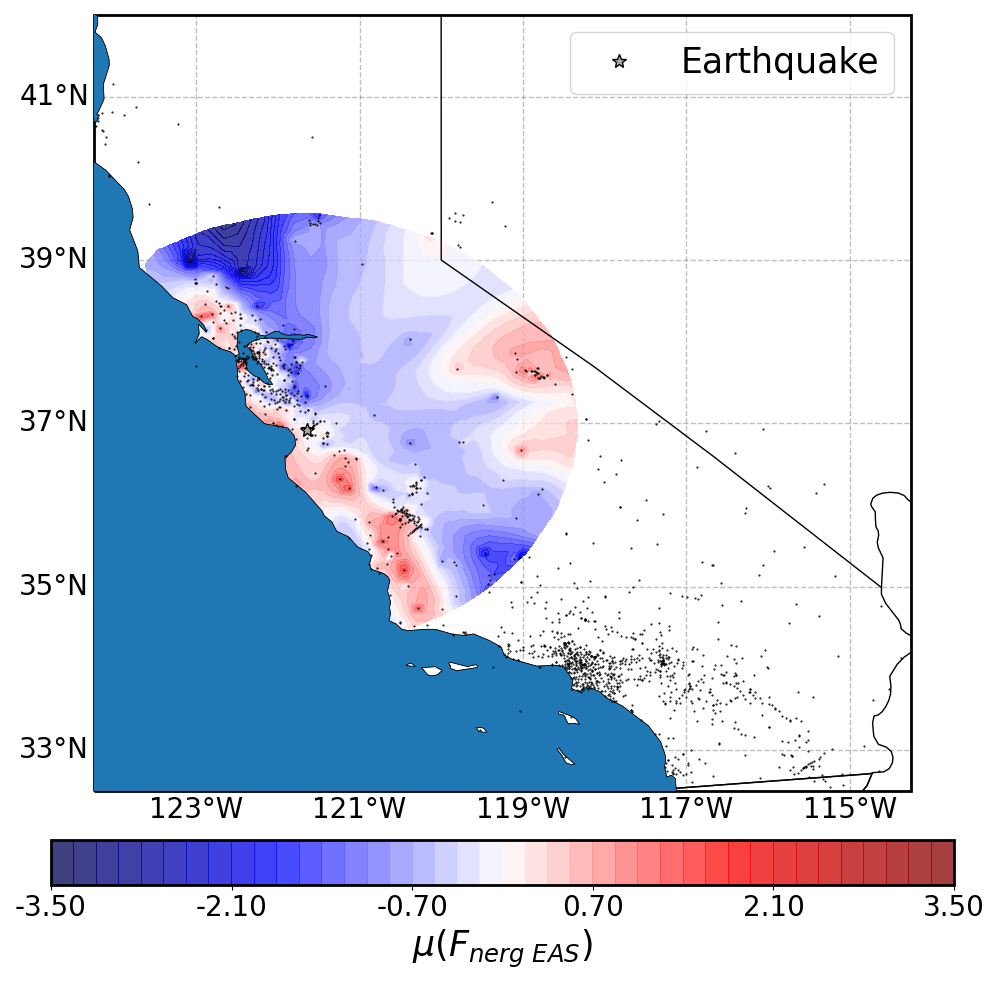}
    \end{subfigure}
    \begin{subfigure}[t]{0.40\textwidth}
        \caption{}  
        \includegraphics[width=.95\textwidth]{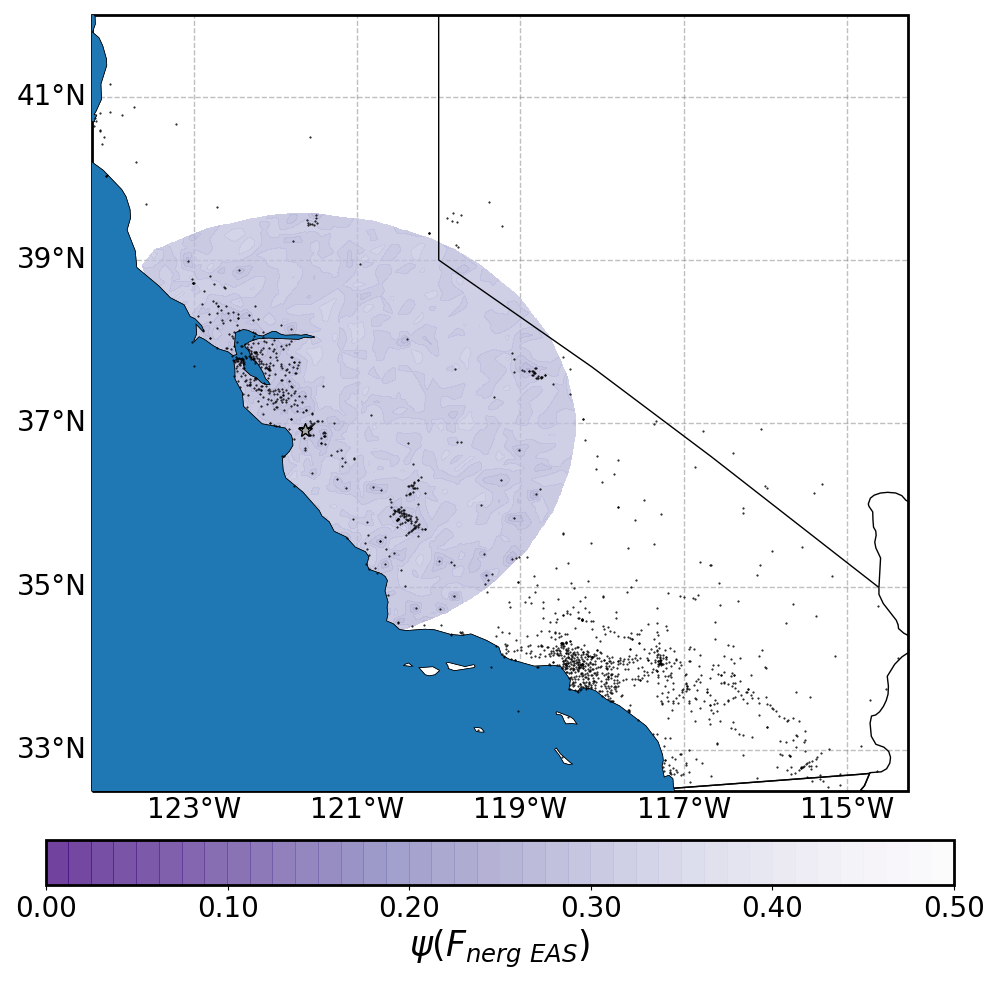}
    \end{subfigure}
    \caption{$EAS$ non-ergodic factors, $F_{nerg~EAS}$, for $f_0 = 10Hz$ for an earthquake in San Andreas. 
    The star corresponds to the earthquake location, and the dots correspond the location of the stations in the used dataset. (a) mean of $F_{nerg~EAS}$ for $M=3.0$, 
    (b) epistemic uncertainty of $F_{nerg~EAS}$ for $M=3.0$
    (c) mean of $F_{nerg~EAS}$ for $M=8.0$, and 
    (d) epistemic uncertainty of $F_{nerg~EAS}$ for $M=8.0$ }
    \label{fig:r_eas_sa_fault}
\end{figure}

\begin{figure}
    \centering
    \begin{subfigure}[t]{0.40\textwidth} 
        \caption{} 
        \includegraphics[width=.95\textwidth]{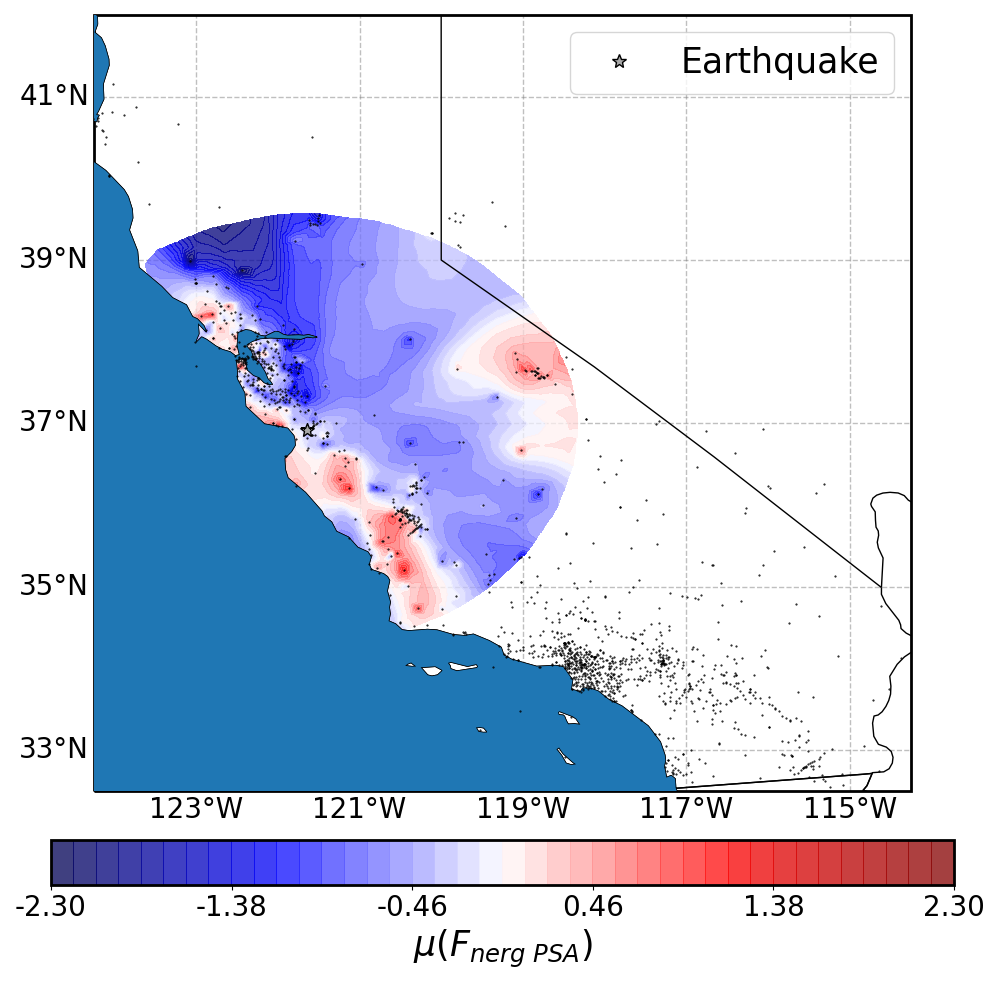}
    \end{subfigure}
    \begin{subfigure}[t]{0.40\textwidth}
        \caption{}  
        \includegraphics[width=.95\textwidth]{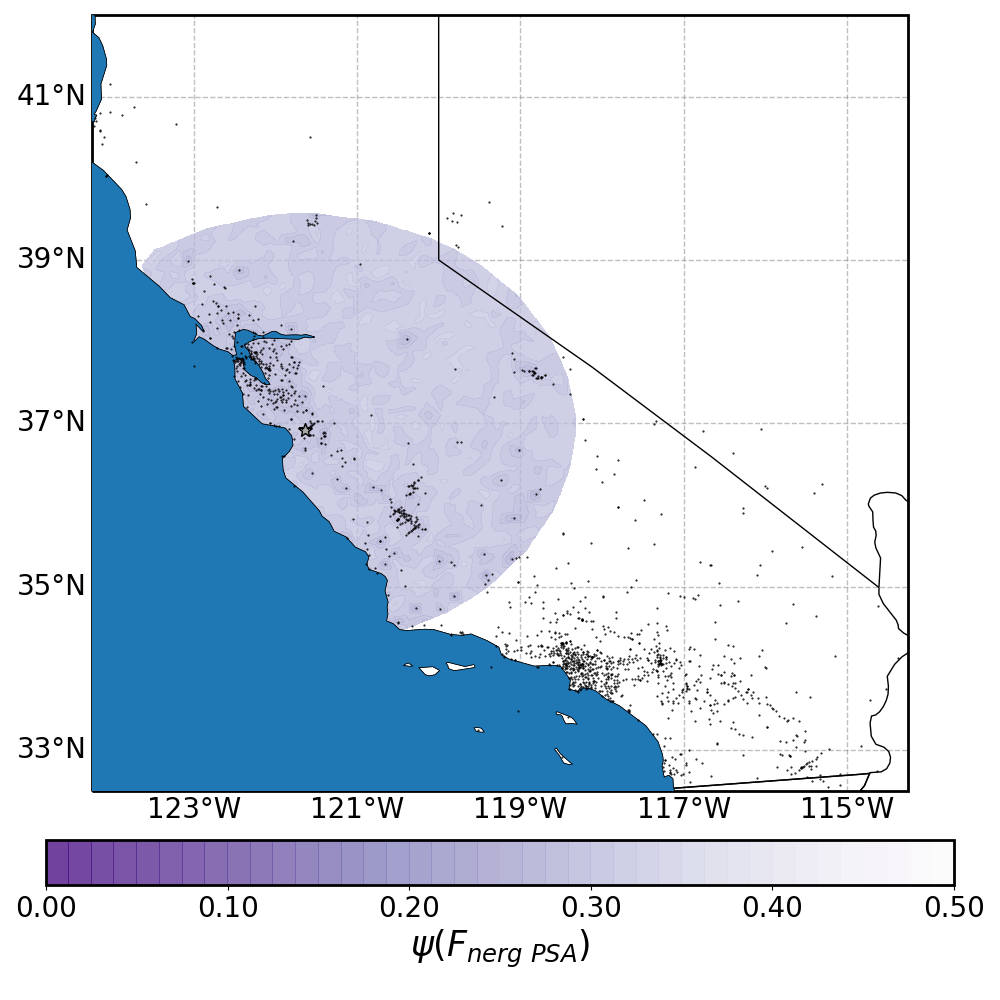}
    \end{subfigure}
    \begin{subfigure}[t]{0.40\textwidth} 
        \caption{} 
        \includegraphics[width=.95\textwidth]{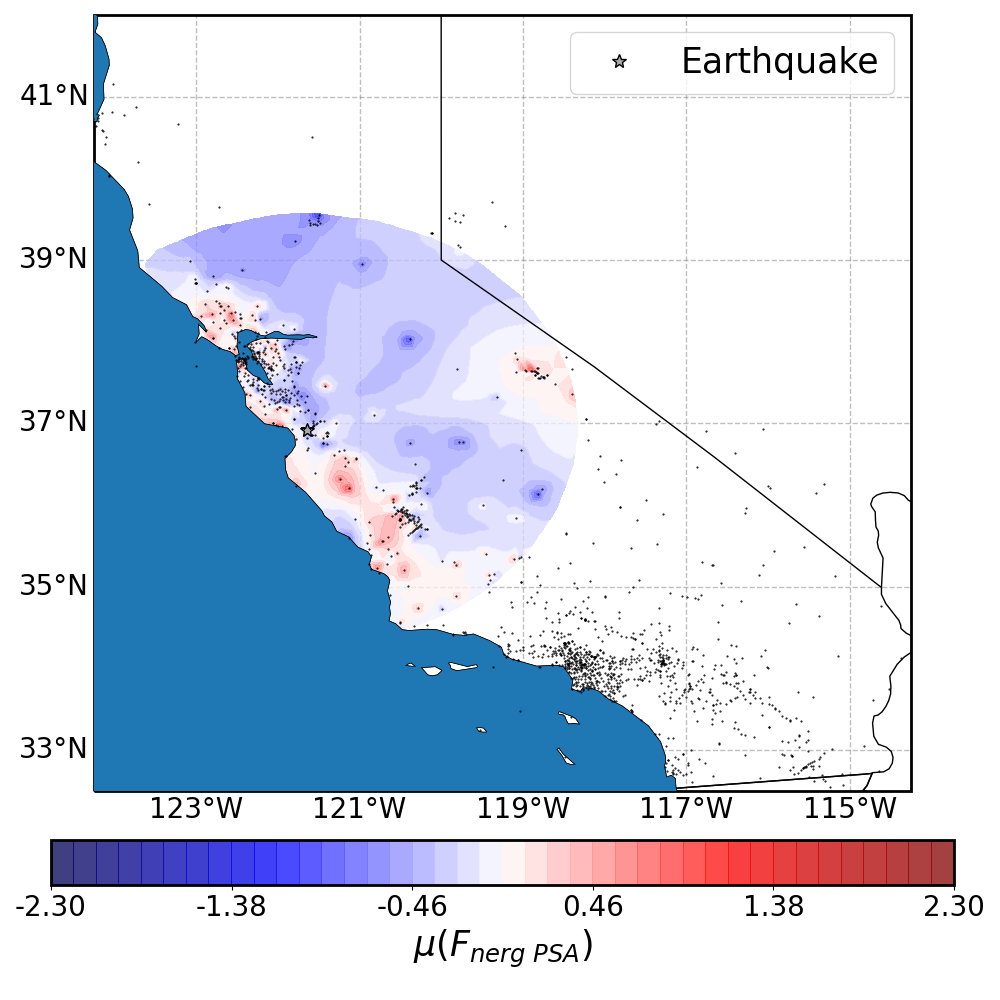}
    \end{subfigure}
    \begin{subfigure}[t]{0.40\textwidth}
        \caption{}  
        \includegraphics[width=.95\textwidth]{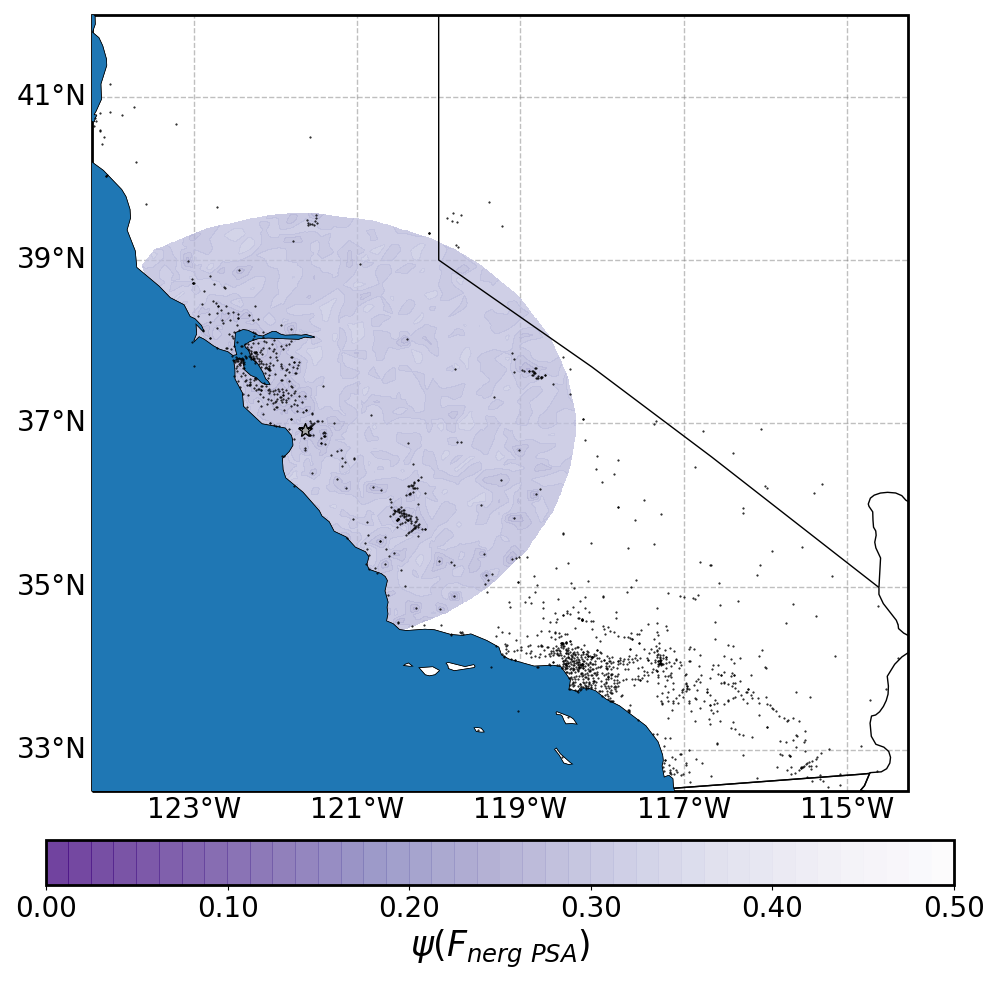}
    \end{subfigure}
    \caption{$PSA$ non-ergodic factors, $F_{nerg~PSA}$, for $T_0 = 0.1sec$ for an earthquake in San Andreas.
    The star corresponds to the earthquake location, and the dots correspond the location of the stations in the used dataset. 
    (a) mean of $F_{nerg~PSA}$ for $M=3.0$, 
    (b) epistemic uncertainty of $F_{nerg~PSA}$ for $M=3.0$
    (c) mean of $F_{nerg~PSA}$ for $M=8.0$, and 
    (d) epistemic uncertainty of $F_{nerg~PSA}$ for $M=8.0$ }
    \label{fig:r_psa_sa_fault}
\end{figure}

As an application example, Figures \ref{fig:r_eas_sa_fault} and \ref{fig:r_psa_sa_fault} present the $EAS$ and $PSA$ non-ergodic for $T_0=0.1sec$ ($f_0 = 10Hz$) for a $M 3$ and $M 8$ earthquake in San Andreas fault. 
The $EAS$ non-ergodic factors are magnitude independent; the median estimate and epistemic uncertainty of $F_{nerg~EAS}$ is the same in both events (Figure \ref{fig:r_eas_sa_fault}).
The magnitude independence allows $F_{nerg EAS}$ to be estimated from the more frequent small magnitude earthquakes and directly applied to the large magnitude events, which are typically of more interest. 
This is not the case for the $PSA$ non-ergodic factors; $F_{nerg PSA}$ depend on the spectral shape; which is why $F_{nerg PSA}$ are different in the $M 3$ and $M 8$ earthquakes (Figure \ref{fig:r_psa_sa_fault}), which illustrates why the non-ergodic $PSA$ GMM is developed with non-ergodic factors that based on $EAS$. 
Most of the regional data that are used to estimate the non-ergodic effects are in form of small magnitude events, which couldn't be used if $PSA$ non-ergodic effects were estimated directly.

In addition, Figures \ref{fig:r_eas_sa_fault} and \ref{fig:r_psa_sa_fault} show the spatial distribution of the epistemic uncertainty.
In this example, where the location of the earthquake is fixed, the spatial distribution of the epistemic uncertainty depends on the path and site location.
Both the $EAS$ and $PSA$ epistemic uncertainties are small near stations that have recorded past events, whereas in remote areas with no available ground-motion data to constrain the non-ergodic terms, the epistemic uncertainties are larger.

\begin{figure}
    \centering
    \begin{subfigure}[t]{0.40\textwidth} 
        \caption{} 
        \includegraphics[width=.95\textwidth]{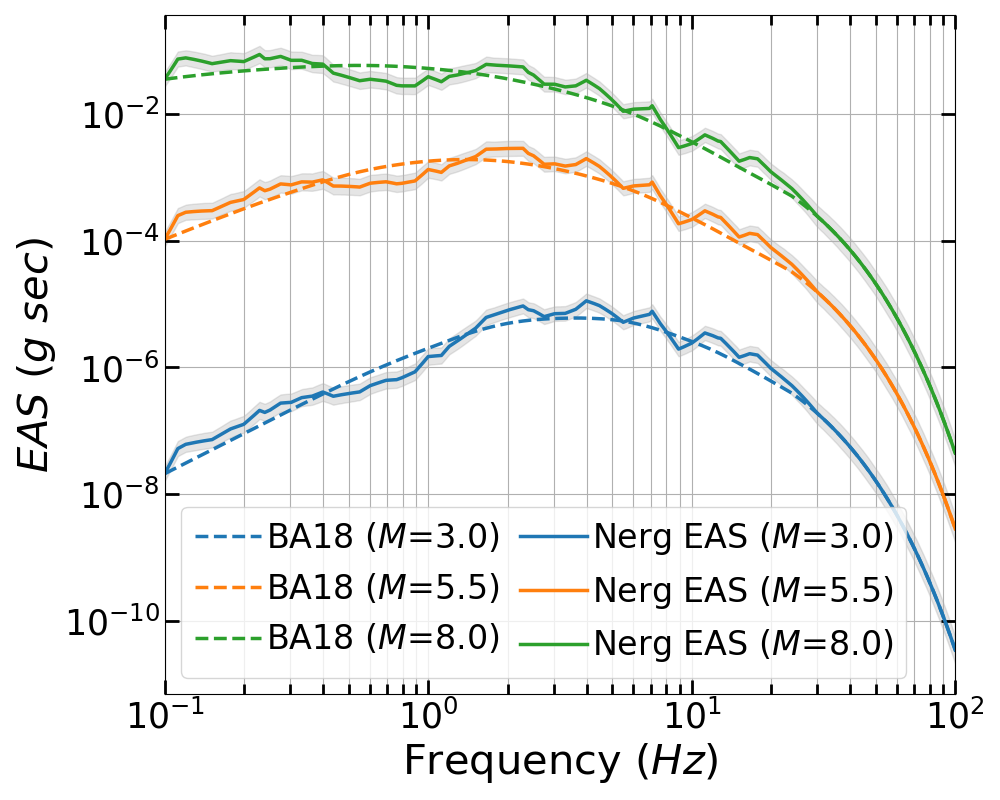}
    \end{subfigure}
    \begin{subfigure}[t]{0.40\textwidth}
        \caption{} 
        \includegraphics[width=.95\textwidth]{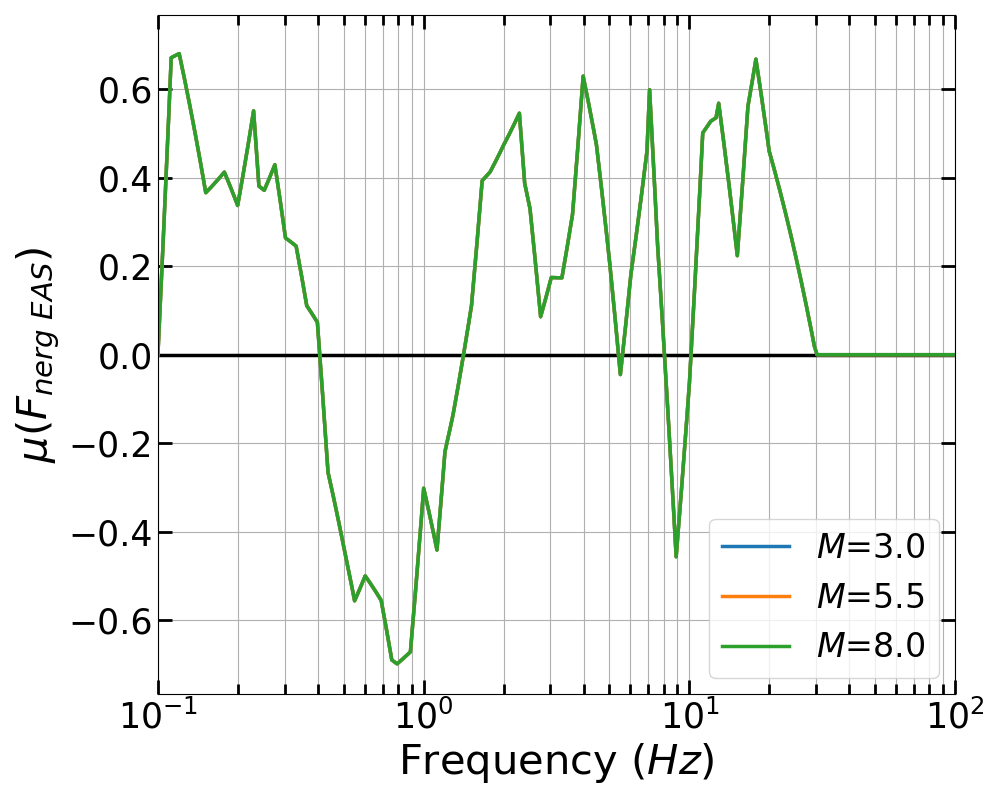}
    \end{subfigure}
    \caption{(a) Ergodic and non-ergodic $EAS$ for $M 3$, $5.5$, and $8$ earthquakes in San Andreas fault, $105km$ from a site in San Francisco, CA 
    (b) non-ergodic $EAS$ factors for the same scenarios.}
    \label{fig:eas_ratios}
\end{figure}

\begin{figure}
    \centering
    \begin{subfigure}[t]{0.40\textwidth} 
        \caption{} 
        \includegraphics[width=.95\textwidth]{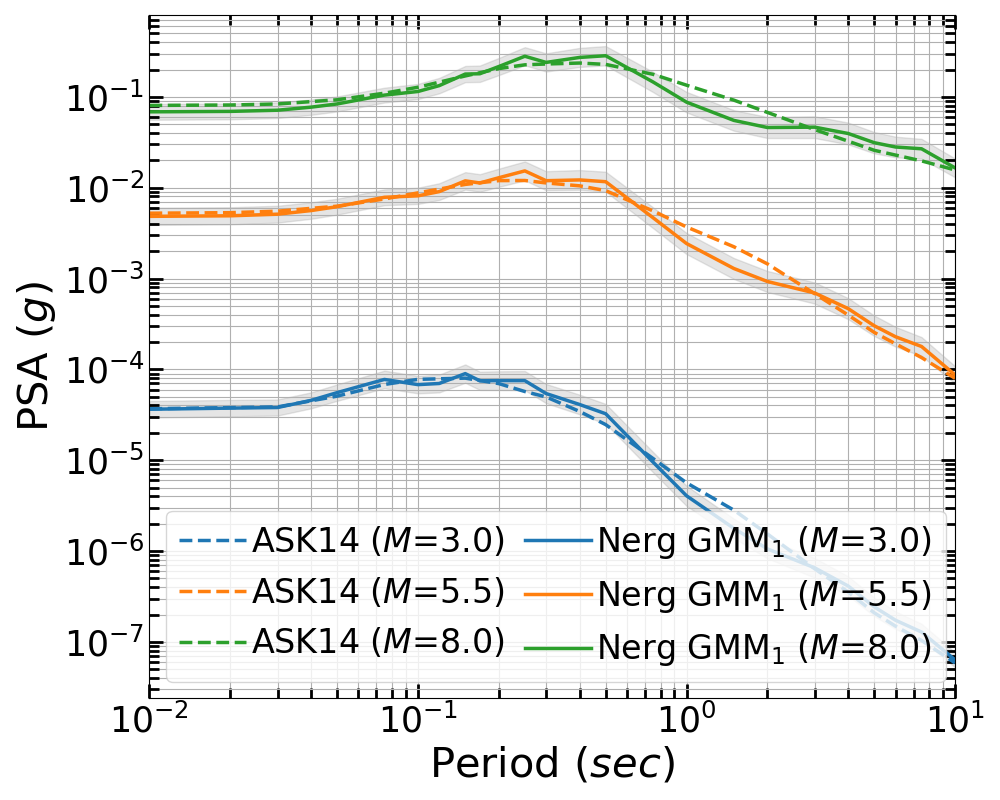}
    \end{subfigure}
    \begin{subfigure}[t]{0.40\textwidth}
        \caption{}  
        \includegraphics[width=.95\textwidth]{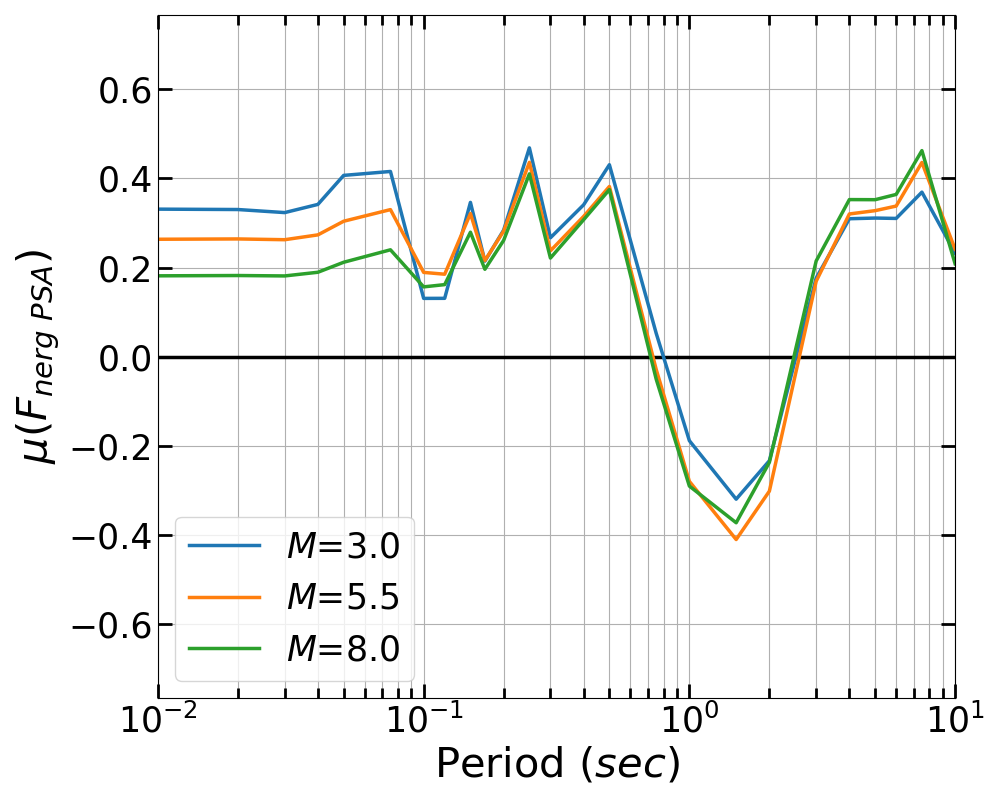}
    \end{subfigure}
    \caption{(a) Ergodic and non-ergodic $PSA$ spectra for $M 3$, $5.5$, and $8$ earthquakes in San Andreas fault, $105km$ from a site in San Francisco, CA 
    (b) non-ergodic $PSA$ factors for the same scenarios.}
    \label{fig:psa_ratios}
\end{figure}

The evaluation of the magnitude dependence of the $EAS$ and $PSA$ non-ergodic factors is further examined in Figures \ref{fig:eas_ratios} and \ref{fig:psa_ratios}.
The three scenarios in this comparison are a $M~3$, $5.5$ and $8$ event in San Andreas Fault, $105 km$ from the site in San Francisco, CA. 
As mentioned previously, the non-ergodic $EAS$ factors are the same for all three events (Figure \ref{fig:eas_ratios}), while the non-ergodic $PSA$ factors are different, especially at small periods (Figure \ref{fig:psa_ratios}), $T_0 < 0.1sec$.
This happens because, for $f_0 > 10Hz$ ($T_0 <0.1sec$), there is little ground-motion content in $EAS$ to resonate the SDOF oscillator, making its response, and subsequently $PSA$, depended on the peak of each spectrum.
Similarly, the non-ergodic $PSA$ factors for $T_0<0.1sec$ depend on the non-ergodic $EAS$ factors at the peak of each spectrum.
In this example, the $M~3$ event has the largest non-ergodic $PSA$ factors at $T_0<0.1sec$, because the non-ergodic $EAS$ factors are predominately positive over its peak ($f=2$ to $6Hz$).
The $M~8$ event has the smallest non-ergodic $PSA$ factors at $T_0<0.1sec$ because its peak ($f<0.1$ to $6Hz$) encompasses the dip of the non-ergodic $EAS$ factors that occur from $f=0.3$ to $2Hz$.

\subsection{Example Hazard Calculations}

\begin{figure}
    \centering
    \begin{subfigure}[t]{0.32\textwidth} 
        \caption{}  \label{fig:haz_erg}
        \includegraphics[width=.95\textwidth]{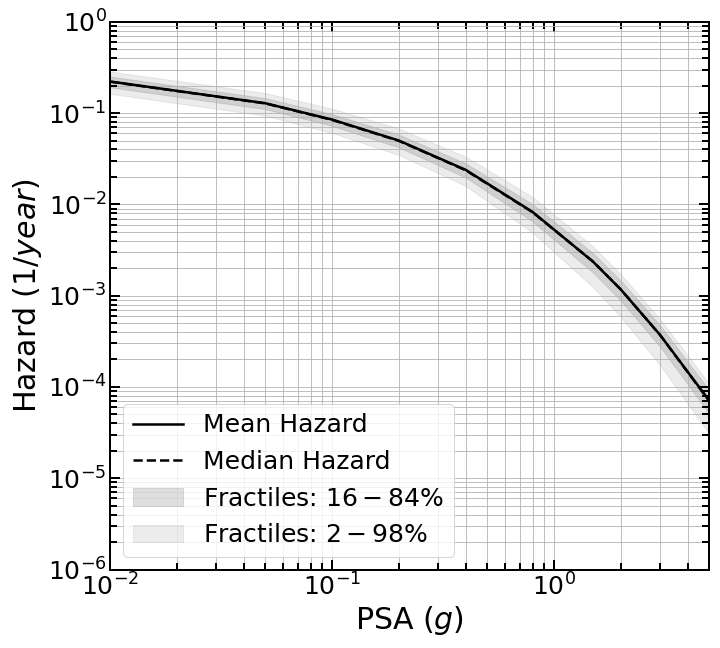}
    \end{subfigure}
    \begin{subfigure}[t]{0.32\textwidth} 
        \caption{} \label{fig:haz_nerg_woSS} 
        \includegraphics[width=.95\textwidth]{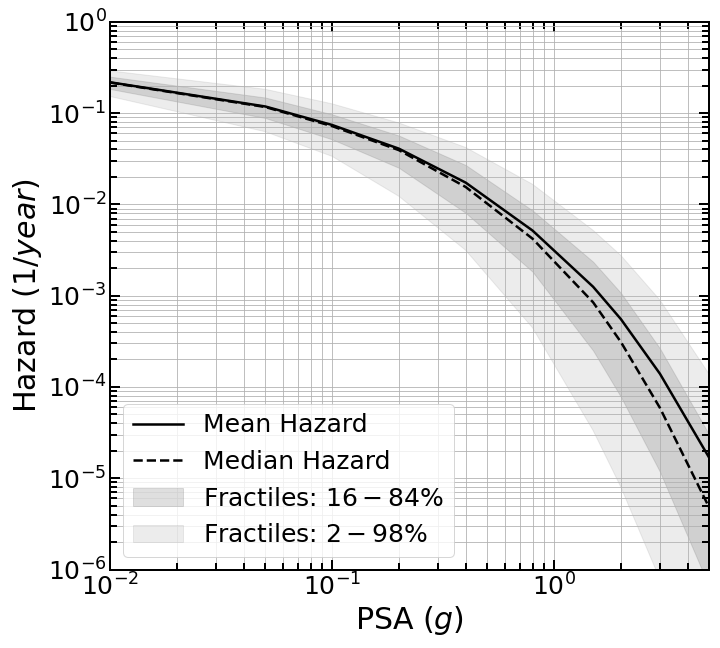}
    \end{subfigure}
    \begin{subfigure}[t]{0.32\textwidth} 
        \caption{}  \label{fig:haz_nerg_wSS}
        \includegraphics[width=.95\textwidth]{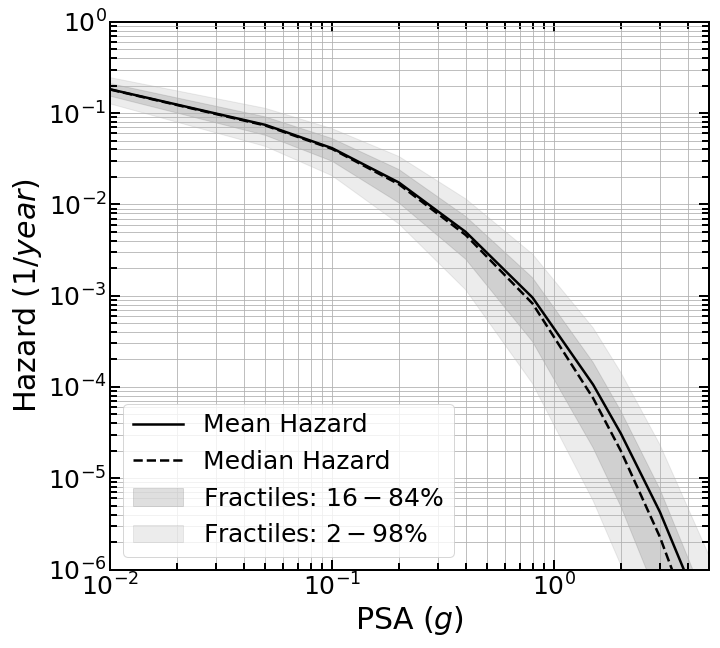}
    \end{subfigure}
    \caption{Hazard results: mean, median, and $2-98\%$, $16-84\%$ fractile ranges of total hazard at $T=0.25sec$ for a site in Berkeley, CA.
    (a) Ergodic hazard calculation,
    (b) Non-ergodic hazard with unconstrained zero-correlation site term ($\delta c_{1b,s}$) of LAK21 GMM, 
    (c) Non-ergodic Hazard with constrained zero-correlation site term ($\delta c_{1b,s}$) of LAK21 GMM.}
    \label{fig:haz}
\end{figure}

A comparison of the ergodic and non-ergodic PSHA results for $PSA(T_0=0.25sec)$ for a site in Berkeley, CA is presented in Figure \ref{fig:haz}.
The PG\&E source model was used in all hazard calculations \citep{PGE2015, PGE2017}. 
The ergodic hazard calculations were performed with the ASK14 and CY14 GMMs, with equal weights, while the non-ergodic hazard calculations were performed with non-ergodic GMM\textsubscript{1} and GMM\textsubscript{2}, with equal weights. 
The epistemic uncertainty of the non-ergodic GMMs was captured by $100$ realizations of $F_{nerg~PSA}$. 
This leads to a logic tree with $200$ branches; each branch is a combination of a non-ergodic model (GMM\textsubscript{1} or GMM\textsubscript{2}) and a $F_{nerg~PSA}$ sample.

The difference between the two non-ergodic hazard calculations is that, in Figure \ref{fig:haz_nerg_woSS}, only the regional systematic site-effects are constrained, while, in Figure \ref{fig:haz_nerg_wSS}, recordings from past earthquakes are assumed to be available and thus, both the regional and site-specific site effects are constrained. 
The regional site effects are captured by the $\delta c_{1a,s}$ term of LAK21 GMM which is a function of the site location.
The site-specific site effects are captured by the $\delta c_{1b,s}$ term of LAK21 GMM, which can be determined either from past recordings or through a site-specific site response analysis. 

For the ergodic hazard calculations, the mean hazard curve is flater than the non-ergodic hazard curves due to the large aleatory variability of ASK14 and CY14, and the epistemic uncertainty is small as it only encompasses the epistemic uncertainty in the seismic source characterization and the median scaling of the ground motion averaged over all of California. 
Comparing the two non-ergodic calculations, the mean hazard curve is flatter and the epistemic uncertainty is larger in Figure \ref{fig:haz_nerg_woSS} as $\delta c_{1b,s}$ is free.
This example shows the impact of non-ergodic GMM in PSHA where at moderate-to-large return periods it can lead to a factor of two to four change in the mean ground-motion level. 

\section{Conclusions} \label{sec:conclusions}

A new approach to develop non-ergodic $PSA$ GMMs is presented in this study which considers the magnitude dependence of the non-ergodic terms.
Due to the linear properties of Fourier Transform, a non-ergodic $EAS$ GMM is used to estimate the non-ergodic effects from the small magnitude events and transfer them to the events of interest.
RVT is used to compute the non-ergodic $PSA$ effects based on the non-ergodic $EAS$ effects, while the average scaling of the non-ergodic $PSA$ GMM is controlled by an existing ergodic $PSA$ GMM.

Two non-ergodic $PSA$ GMMs are developed in this study.
The first one uses the ASK14 GMM as a backbone model for the average scaling and is applicable to periods $T_0 = 0.01 - 10 sec$.
The second one uses the CY14 GMM as a backbone model for the average scaling and is applicable to periods $T_0 = 0.01 - 5 sec$.
The non-ergodic $PSA$ effects are quantified in terms of non-erodic $PSA$ factors, that is the difference between the log of $PSA$ estimated with RVT and the non-ergodic $EAS$ and the log of $PSA$ estimated with RVT and the ergodic $EAS$. 
In both cases, the LAK21 GMM is used for the non-ergodic $EAS$ and the BA18 GMM is used for the ergodic $EAS$. 
The RVT calculations are performed with the V75 $PF$, the median estimate of $D_{a5-85}$ from AS96 for the ground-motion duration, and the BT15 for the oscillator duration. 
The RVT components were chosen based on a thorough evaluation of alternative models for the peak factors, ground-motion duration and oscillator duration.
The objective of the evaluation was to minimize misfit between the observed $PSA$ and the $PSA$ computed with RVT. 

The advantages of developing the non-ergodic GMM with an ergodic backbone model and non-ergodic $PSA$ factors, instead of developing it directly with RVT and the LAK21 are: i) the elimination of the small bias of $RVT$ at $T_0 = 1 - 4sec$, ii) the separation of the non-ergodic effects from average scaling, and iii) the adoption of complex scaling terms present in ergodic $PSA$ GMMs.
Compared to the recorded $PSA$, the $PSA$ estimated with RVT has a small positive bias at $T_0 = 1 - 4sec$.
This bias is not propagated in the non-ergodic $PSA$ factors; it is canceled out, as both the ergodic and non-ergodic RVT $PSA$ estimates are calculated with the same approach.

Aleatory aleatory variability of the two non-ergodic $PSA$ GMMs is approximately $30$ to $35\%$ smaller than the aleatory variability of an ergodic $PSA$ GMM.

Future studies should reevaluate the RVT and $EAS$ models so that when combined they result in a $PSA$ predictions consistent with $PSA$ GMMs. 
Furthermore, the proposed non-ergodic GMMs were developed with a subset of the NGAWest2 database which was compiled in 2014.
As larger data sets which include more recent and more frequent small magnitude events become available, the proposed models should be assessed and potentially expanded with additional non-ergodic terms.
Similarly, $3D$ broadband numerical simulations or inferred intensity measurements from historical earthquakes should be used to evaluate the efficacy of the proposed models. 

\section{Software and Resources}
The RVT calculations were performed with the pyRVT library \citep{pyrvt} in the computer language Python \citep{python}. 
The linear mixed-effects regressions were performed with the lme4 package \citep{lme4} in the statistical environment R \citep{R}.
The PSHA calculations were performed with HAZ45.3 \citep{HAZ45}.

\section{Acknowledgements}
This work was partially supported by the PG\&E Geosciences Department Long-Term Seismic Program. 
The authors thank Nicolas Kuehn for instructive comments on an early draft of this manuscript.

\section*{Declarations}
\subsection*{Funding}
This work was partially funded by the PG\&E Geosciences Department Long-Term Seismic Program.

\subsection*{Conflict of interest}
The authors declare that they have no conflict of interest.

\subsection*{Ethics approval}
Non applicable

\subsection*{Consent to participate}
Non applicable 

\subsection*{Consent for publication}
Non applicable 

\subsection*{Availability of data and material}

\subsection*{Code availability}
The are python scripts for the non-ergodic regressions are provided at:\\
{\fontfamily{qcr}\selectfont
https://github.com/glavrentiadis/NonErgodicGMM\_public
}

\bibliographystyle{chicago}
\bibliography{references_mendeley_GL.bib, references_other.bib}

\end{document}


\maketitle

\section{Comparison of different duration intervals for $D_{gm}$} \label{esup:sec:rvt_dur_gm}

\begin{figure}[H]
    \centering
    \begin{subfigure}[t]{0.40\textwidth} 
        \caption{} 
        \includegraphics[width=.95\textwidth]{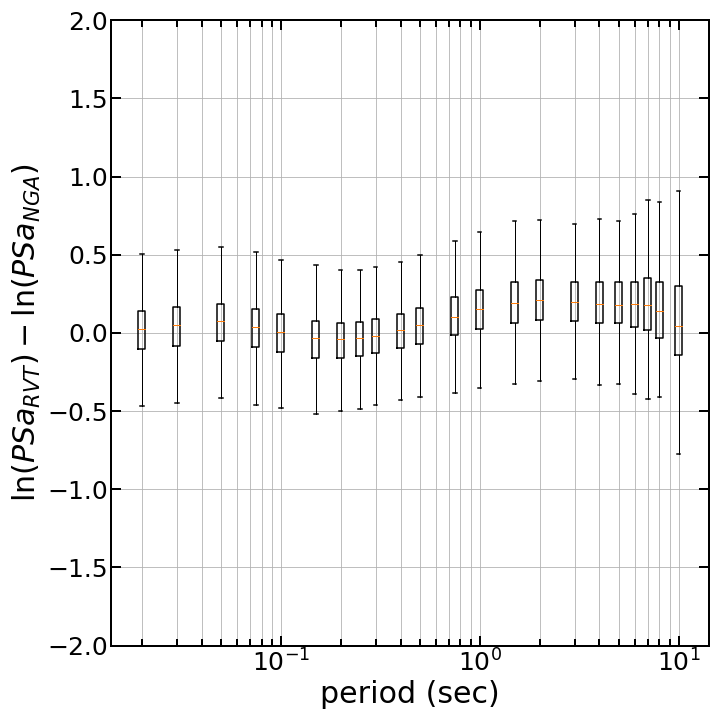}
    \end{subfigure}
    \begin{subfigure}[t]{0.40\textwidth} 
        \caption{} 
        \includegraphics[width=.95\textwidth]{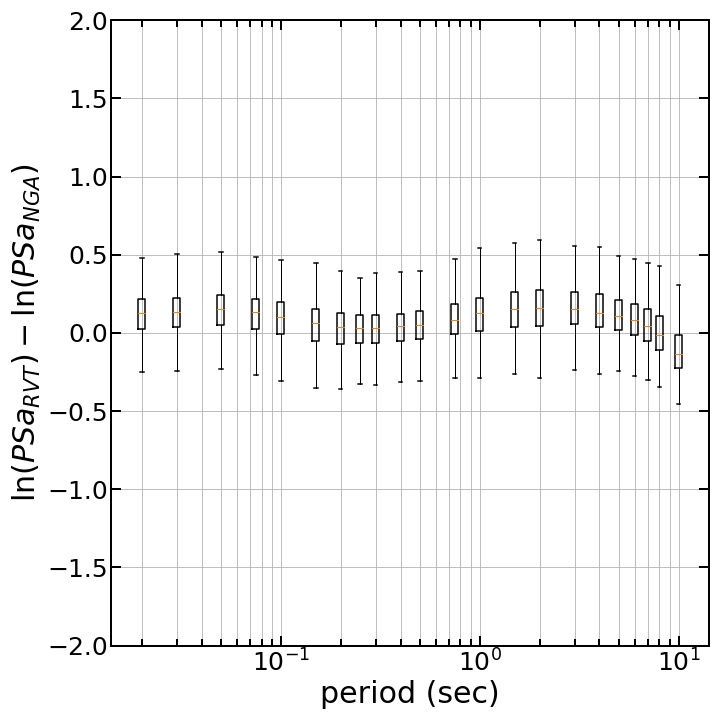}
    \end{subfigure}
    \caption{Residuals between the records' $PSa$ and the RVT $PSa$. V75 $PF$, records' $D_{a0.05-0.75}$ as $D_{gm}$, and BT15 $D_{rms}$. (a) residuals for all $M$, (b) residuals for $M > 5$ }
    \label{esup:fig:cmp_PF_BT15_dur_actual_d0.05-0.75}
\end{figure}

\begin{figure}[H]
    \centering
    \begin{subfigure}[t]{0.40\textwidth} 
        \caption{} 
        \includegraphics[width=.95\textwidth]{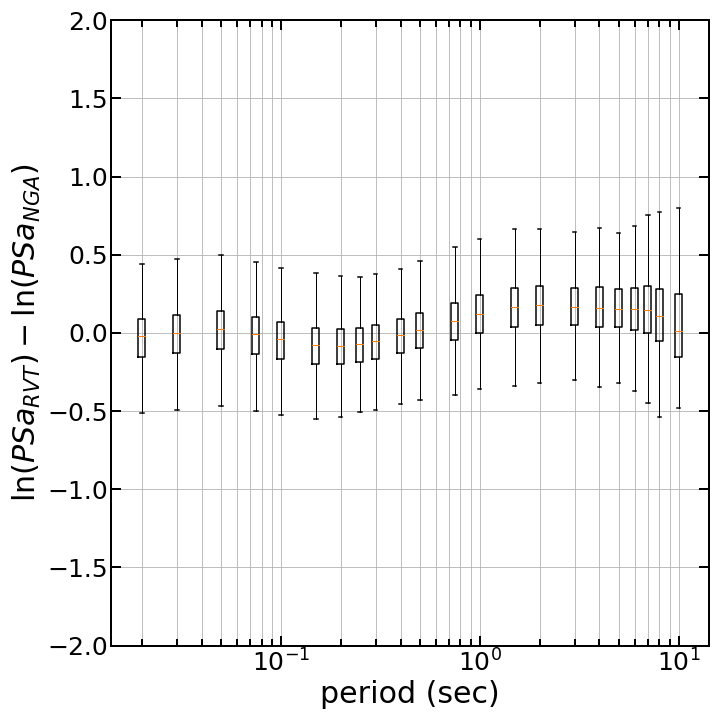}
    \end{subfigure}
    \begin{subfigure}[t]{0.40\textwidth} 
        \caption{} 
        \includegraphics[width=.95\textwidth]{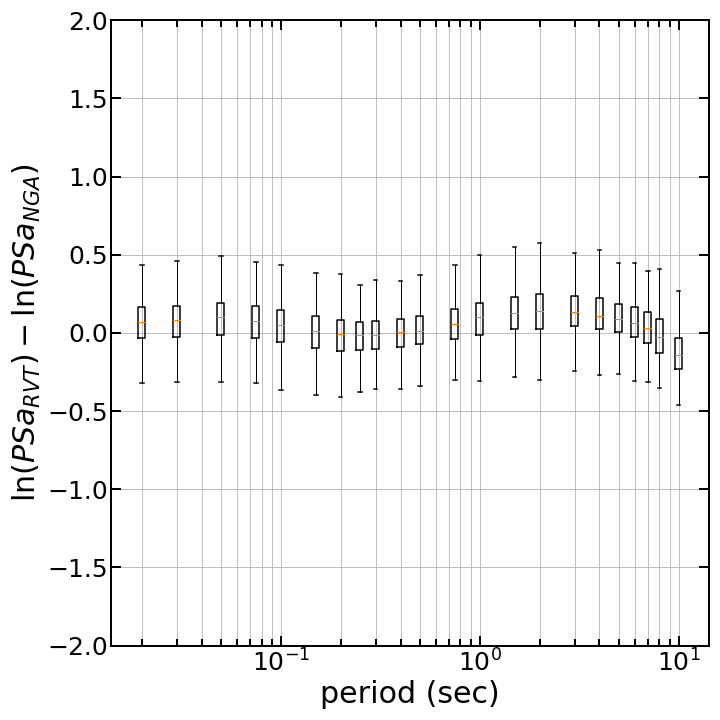}
    \end{subfigure}
    \caption{Residuals between the records' $PSa$ and the RVT $PSa$. V75 $PF$, records' $D_{a0.05-0.80}$ as $D_{gm}$, and BT15 $D_{rms}$. (a) residuals for all $M$, (b) residuals for $M > 5$ }
    \label{esup:fig:cmp_PF_BT15_dur_actual_d0.05-0.80}
\end{figure}

\begin{figure}[H]
    \centering
    \begin{subfigure}[t]{0.40\textwidth} 
        \caption{} 
        \includegraphics[width=.95\textwidth]{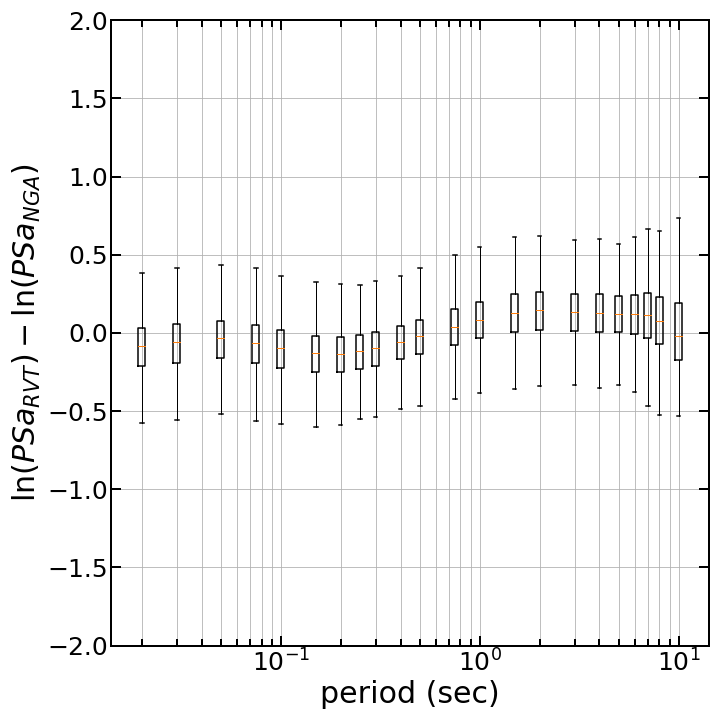}
    \end{subfigure}
    \begin{subfigure}[t]{0.40\textwidth} 
        \caption{} 
        \includegraphics[width=.95\textwidth]{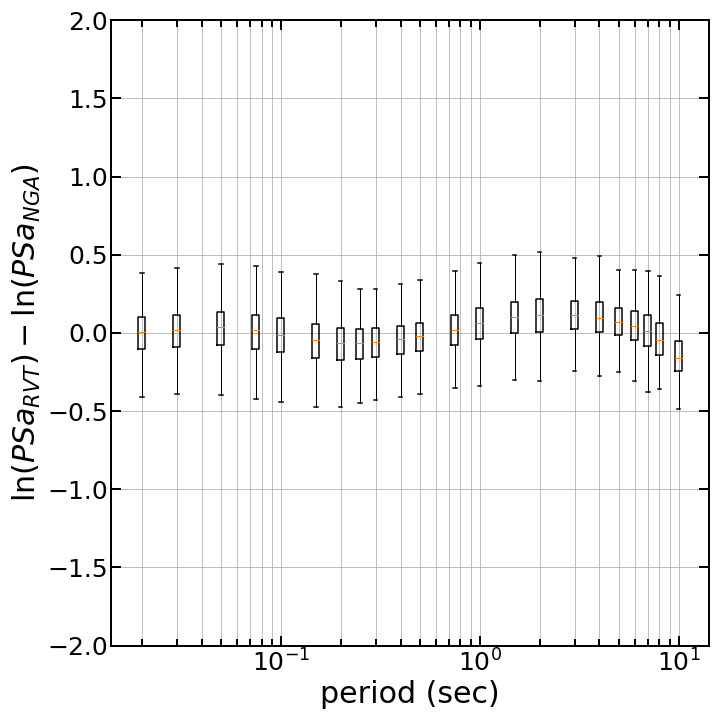}
    \end{subfigure}
    \caption{Residuals between the records' $PSa$ and the RVT $PSa$. V75 $PF$, records' $D_{a0.05-0.85}$ as $D_{gm}$, and BT15 $D_{rms}$. (a) residuals for all $M$, (b) residuals for $M > 5$ }
    \label{esup:fig:cmp_PF_BT15_dur_actual_d0.05-0.85}
\end{figure}

\begin{figure}[H]
    \centering
    \begin{subfigure}[t]{0.40\textwidth} 
        \caption{} 
        \includegraphics[width=.95\textwidth]{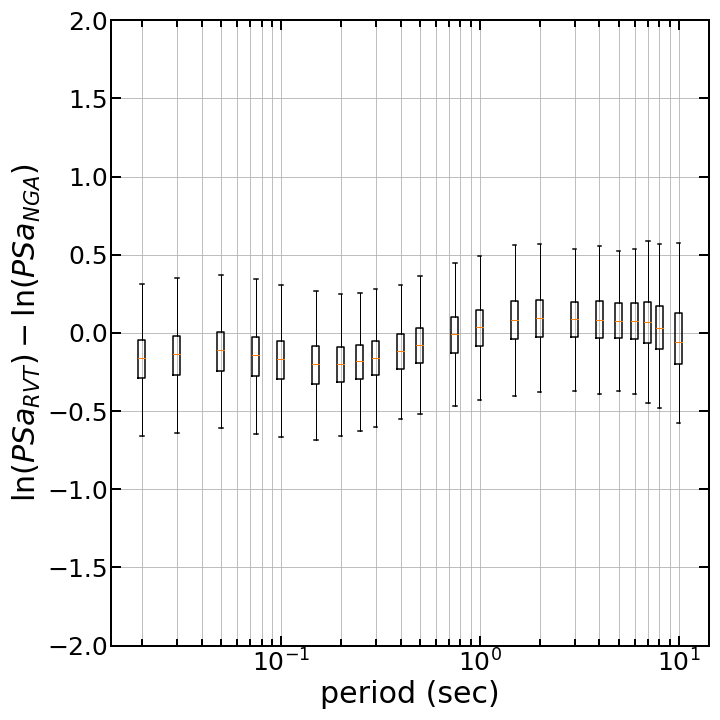}
    \end{subfigure}
    \begin{subfigure}[t]{0.40\textwidth} 
        \caption{} 
        \includegraphics[width=.95\textwidth]{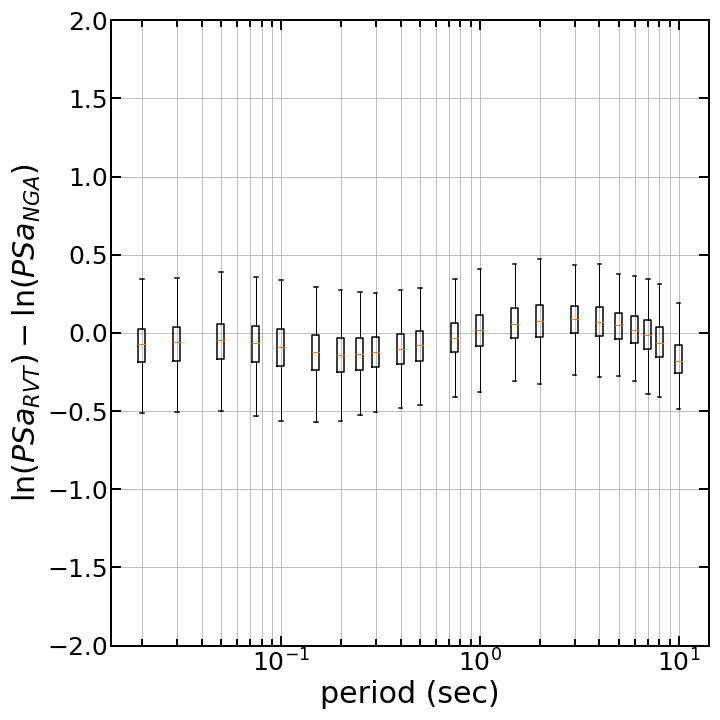}
    \end{subfigure}
    \caption{Residuals between the records' $PSa$ and the RVT $PSa$. V75 $PF$, records' $D_{a0.05-0.90}$ as $D_{gm}$, and BT15 $D_{rms}$. (a) residuals for all $M$, (b) residuals for $M > 5$ }
    \label{esup:fig:cmp_PF_BT15_dur_actual_d0.05-0.90}
\end{figure}

\begin{figure}[H]
    \centering
    \begin{subfigure}[t]{0.40\textwidth} 
        \caption{} 
        \includegraphics[width=.95\textwidth]{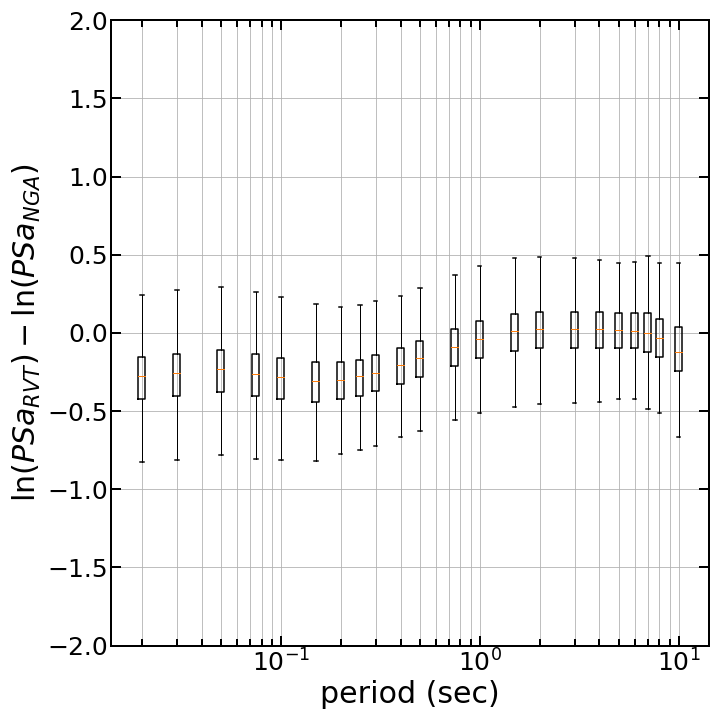}
    \end{subfigure}
    \begin{subfigure}[t]{0.40\textwidth} 
        \caption{} 
        \includegraphics[width=.95\textwidth]{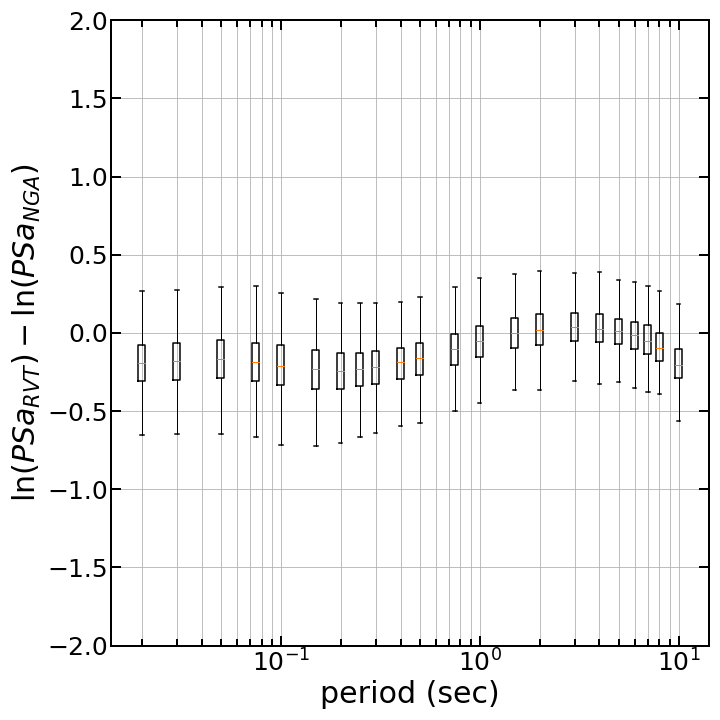}
    \end{subfigure}
    \caption{Residuals between the records' $PSa$ and the RVT $PSa$. V75 $PF$, records' $D_{a0.05-0.95}$ as $D_{gm}$, and BT15 $D_{rms}$. (a) residuals for all $M$, (b) residuals for $M > 5$ }
    \label{esup:fig:cmp_PF_BT15_dur_actual_d0.05-0.95}
\end{figure}

Figures \ref{esup:fig:cmp_PF_BT15_dur_actual_d0.05-0.75} to \ref{esup:fig:cmp_PF_BT15_dur_actual_d0.05-0.95} show the residuals between the records' $PSa$ and the $PSa$ estimated with RVT. The RVT $PSa$ were estimated with $V75$ peak factors, records' actual duration for $D_{gm}$, and $BT15$ for $D_{rms}$.
$D_{gm}$ is equal to: $D_{a5-75}$ in Figure \ref{esup:fig:cmp_PF_BT15_dur_actual_d0.05-0.75}, $D_{a5-80}$ in Figure \ref{esup:fig:cmp_PF_BT15_dur_actual_d0.05-0.80}, $D_{a5-85}$ in Figure \ref{esup:fig:cmp_PF_BT15_dur_actual_d0.05-0.85}, $D_{a5-90}$ in Figure \ref{esup:fig:cmp_PF_BT15_dur_actual_d0.05-0.90}, and $D_{a5-95}$ in Figure \ref{esup:fig:cmp_PF_BT15_dur_actual_d0.05-0.95}.

\newpage
\section{Comparison of different duration models for $D_{gm}$} \label{esup:sec:rvt_dur_gmm}

\begin{figure}[H]
    \centering
    \begin{subfigure}[t]{0.40\textwidth} 
        \caption{} 
        \includegraphics[width=.95\textwidth]{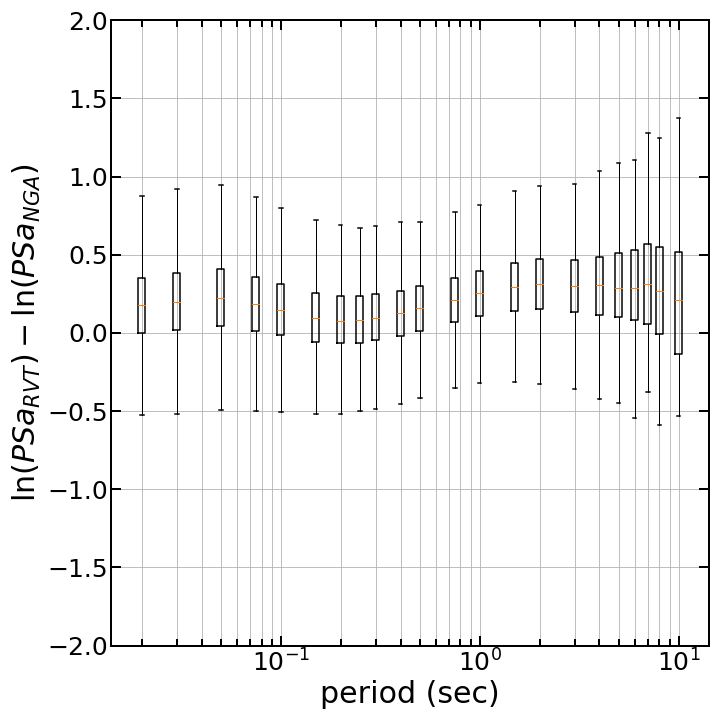}
    \end{subfigure}
    \begin{subfigure}[t]{0.40\textwidth} 
        \caption{} 
        \includegraphics[width=.95\textwidth]{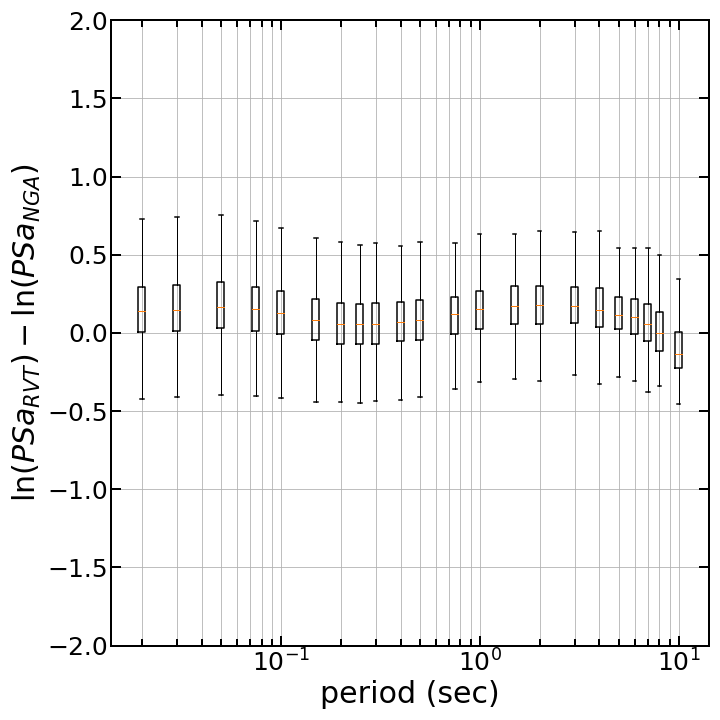}
    \end{subfigure}
    \caption{Residuals between the records' $PSa$ and the RVT $PSa$. V75 $PF$, $D_{a0.05-0.75}$ from KS06 as $D_{gm}$, and BT15 $D_{rms}$. (a) residuals for all $M$, (b) residuals for $M > 5$}
    \label{esup:fig:cmp_PF_BT15_dur_KS06_da0.05-0.75}
\end{figure}

\begin{figure}[H]
    \centering
    \begin{subfigure}[t]{0.40\textwidth} 
        \caption{} 
        \includegraphics[width=.95\textwidth]{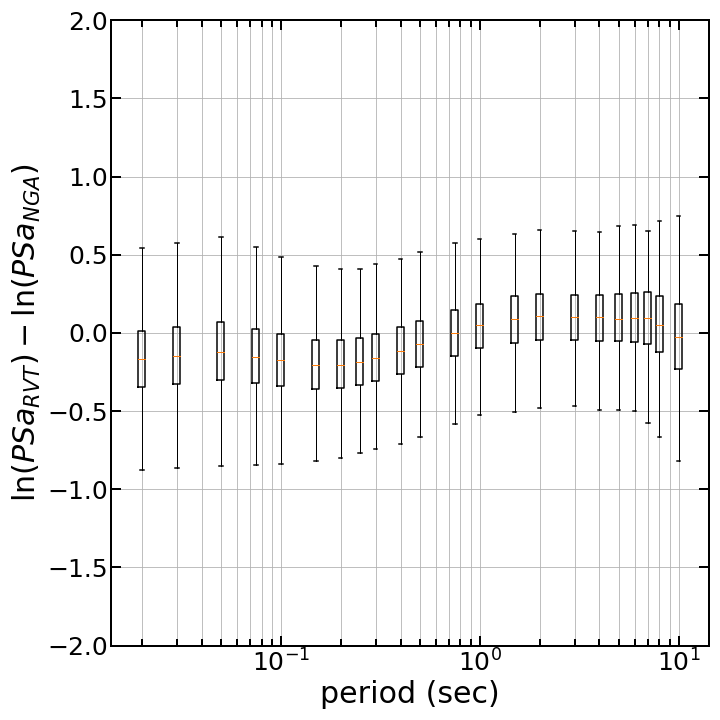}
    \end{subfigure}
    \begin{subfigure}[t]{0.40\textwidth} 
        \caption{} 
        \includegraphics[width=.95\textwidth]{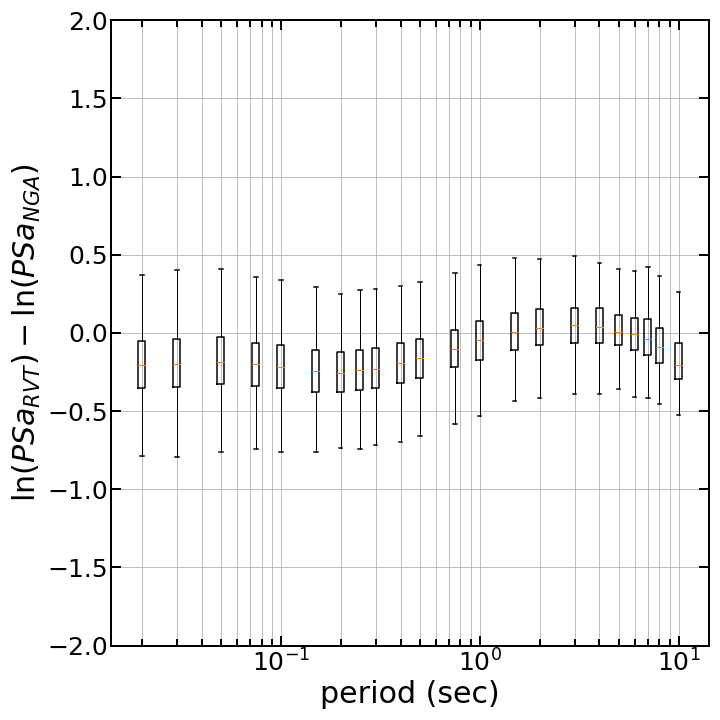}
    \end{subfigure}
    \caption{Residuals between the records' $PSa$ and the RVT $PSa$. V75 $PF$, $D_{a0.05-0.95}$ from KS06 as $D_{gm}$, and BT15 $D_{rms}$. (a) residuals for all $M$, (b) residuals for $M > 5$}
    \label{esup:fig:cmp_PF_BT15_dur_KS06_da0.05-0.95}
\end{figure}

\begin{figure}[H]
    \centering
    \begin{subfigure}[t]{0.40\textwidth} 
        \caption{} 
        \includegraphics[width=.95\textwidth]{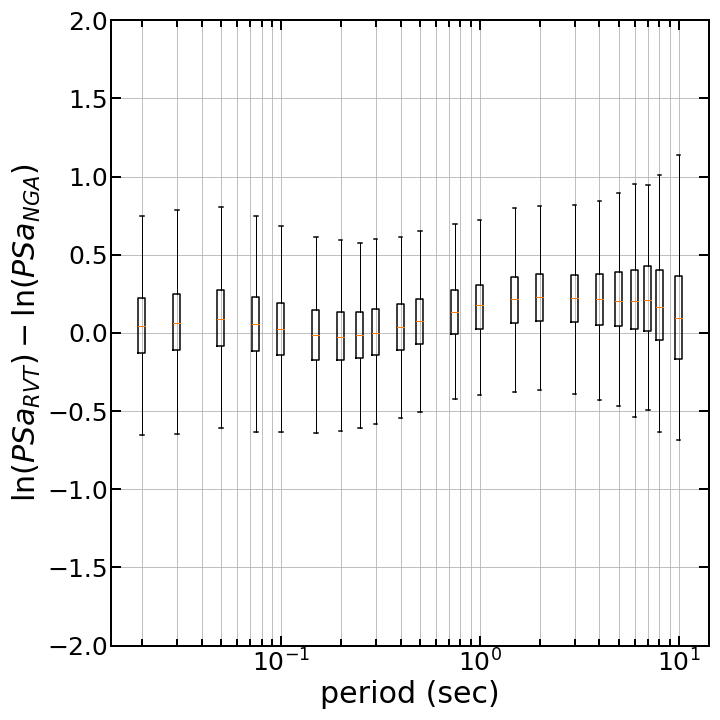}
    \end{subfigure}
    \begin{subfigure}[t]{0.40\textwidth} 
        \caption{} 
        \includegraphics[width=.95\textwidth]{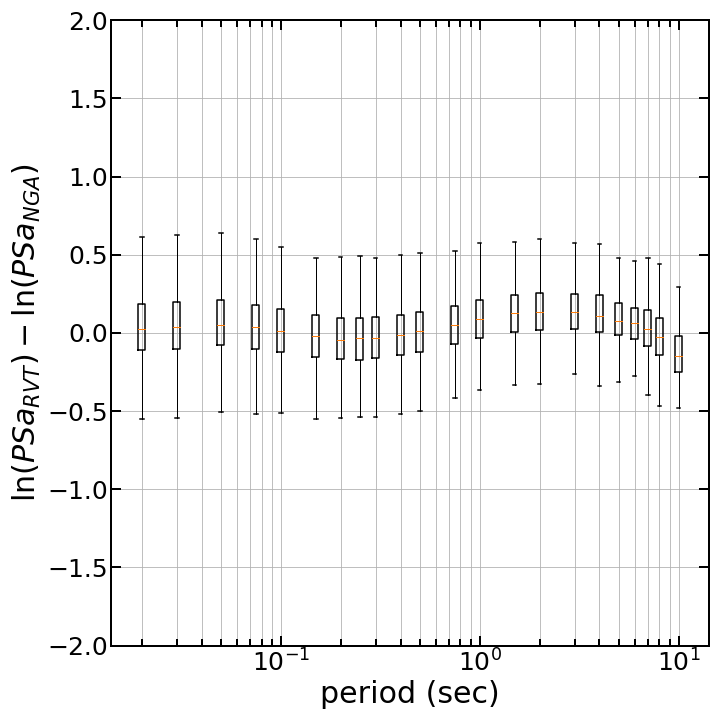}
    \end{subfigure}
    \caption{Residuals between the records' $PSa$ and the RVT $PSa$. V75 $PF$, $D_{v0.05-0.75}$ from KS06 as $D_{gm}$, and BT15 $D_{rms}$. (a) residuals for all $M$, (b) residuals for $M > 5$}
    \label{esup:fig:cmp_PF_BT15_dur_KS06_dv0.05-0.75}
\end{figure}

\begin{figure}[H]
    \centering
    \begin{subfigure}[t]{0.40\textwidth} 
        \caption{} 
        \includegraphics[width=.95\textwidth]{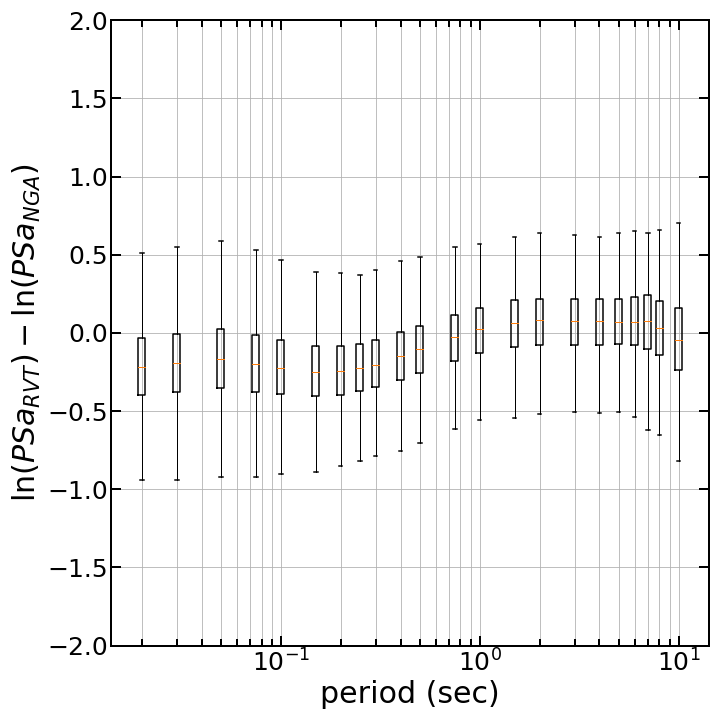}
    \end{subfigure}
    \begin{subfigure}[t]{0.40\textwidth} 
        \caption{} 
        \includegraphics[width=.95\textwidth]{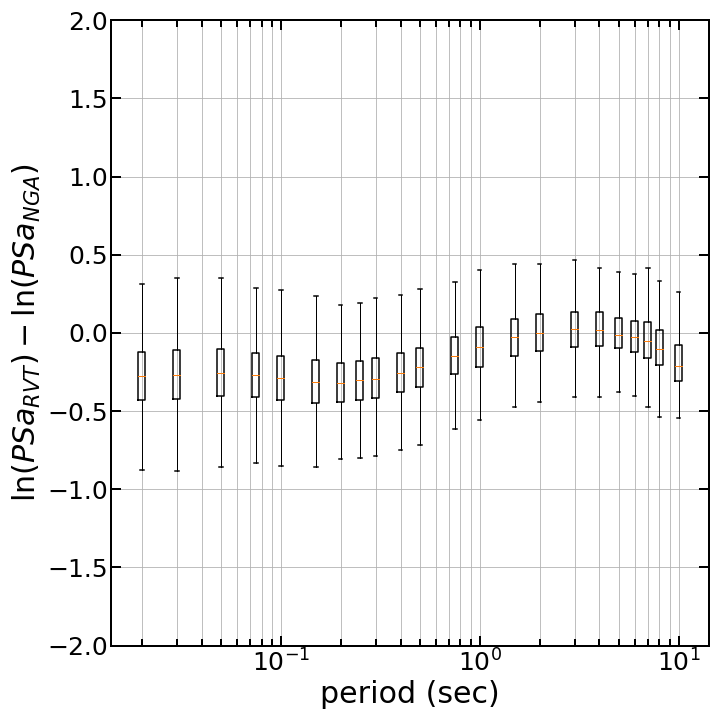}
    \end{subfigure}
    \caption{Residuals between the records' $PSa$ and the RVT $PSa$. V75 $PF$, $D_{v0.05-0.95}$ from KS06 as $D_{gm}$, and BT15 $D_{rms}$. (a) residuals for all $M$, (b) residuals for $M > 5$}
    \label{esup:fig:cmp_PF_BT15_dur_KS06_dv0.05-0.95}
\end{figure}

\begin{figure}[H]
    \centering
    \begin{subfigure}[t]{0.40\textwidth} 
        \caption{} 
        \includegraphics[width=.95\textwidth]{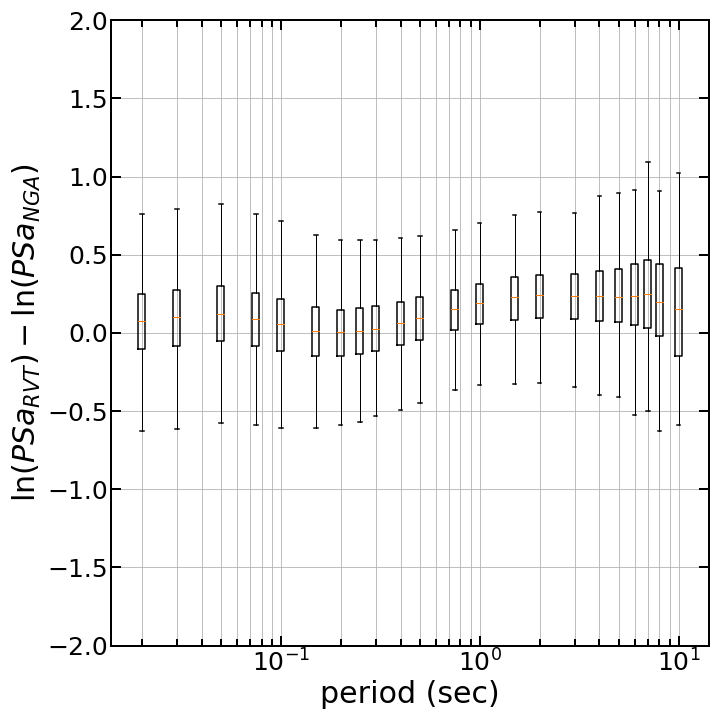}
    \end{subfigure}
    \begin{subfigure}[t]{0.40\textwidth} 
        \caption{} 
        \includegraphics[width=.95\textwidth]{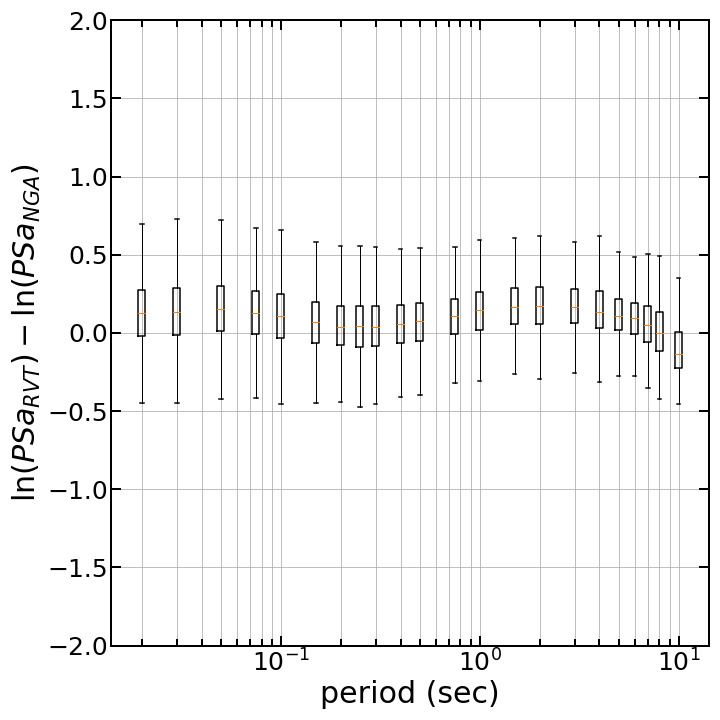}
    \end{subfigure}
    \caption{Residuals between the records' $PSa$ and the RVT $PSa$. V75 $PF$, $D_{a0.05-0.75}$ from KS06 as $D_{gm}$, and BT15 $D_{rms}$. (a) residuals for all $M$, (b) residuals for $M > 5$}
    \label{esup:fig:cmp_PF_BT15_dur_AS16_da0.05-0.75}
\end{figure}

\begin{figure}[H]
    \centering
    \begin{subfigure}[t]{0.40\textwidth} 
        \caption{} 
        \includegraphics[width=.95\textwidth]{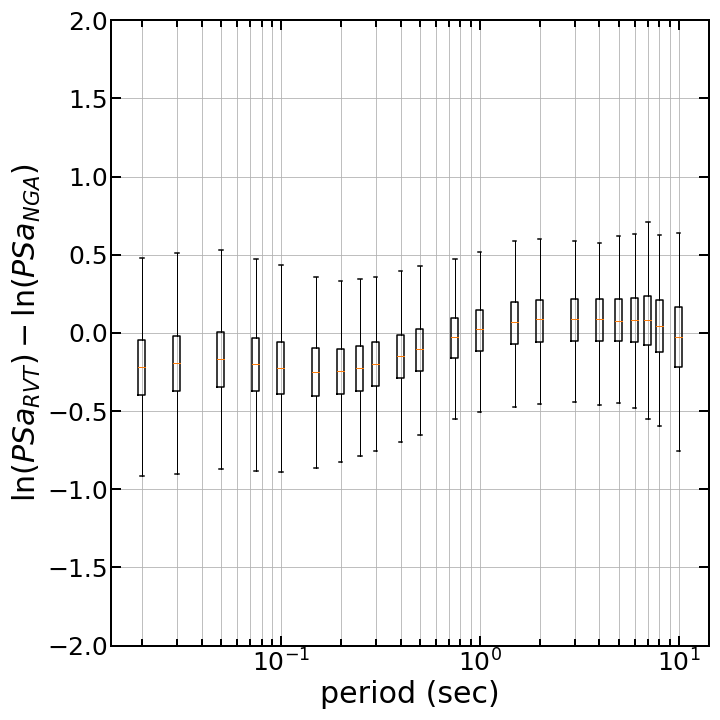}
    \end{subfigure}
    \begin{subfigure}[t]{0.40\textwidth} 
        \caption{} 
        \includegraphics[width=.95\textwidth]{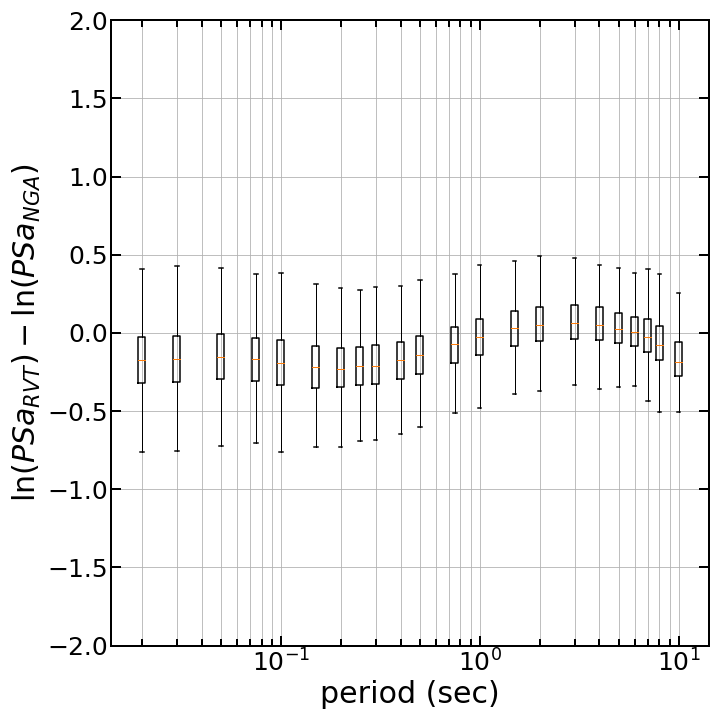}
    \end{subfigure}
    \caption{Residuals between the records' $PSa$ and the RVT $PSa$. V75 $PF$, $D_{a0.05-0.95}$ from KS06 as $D_{gm}$, and BT15 $D_{rms}$. (a) residuals for all $M$, (b) residuals for $M > 5$}
    \label{esup:fig:cmp_PF_BT15_dur_AS16_da0.05-0.95}
\end{figure}

\begin{figure}[h]
    \centering
    \begin{subfigure}[t]{0.40\textwidth} 
        \caption{} 
        \includegraphics[width=.95\textwidth]{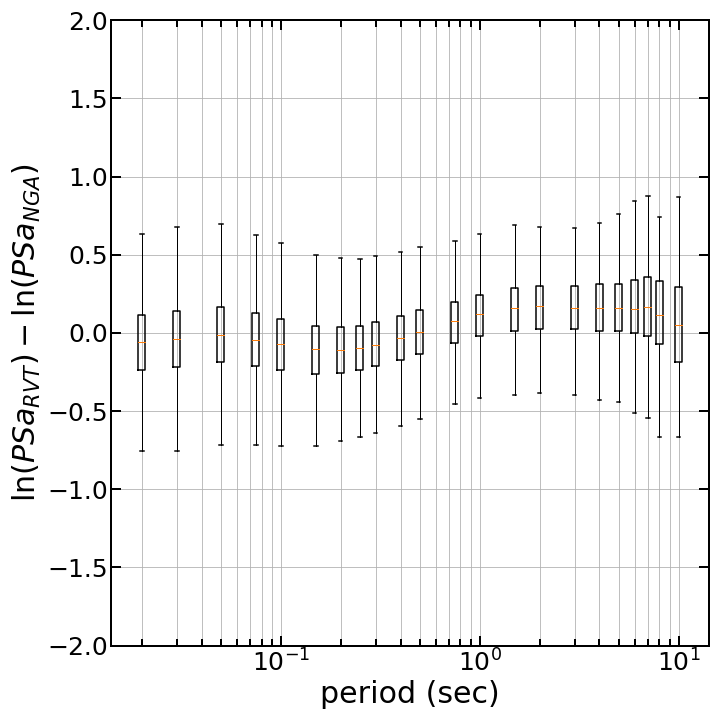}
    \end{subfigure}
    \begin{subfigure}[t]{0.40\textwidth} 
        \caption{} 
        \includegraphics[width=.95\textwidth]{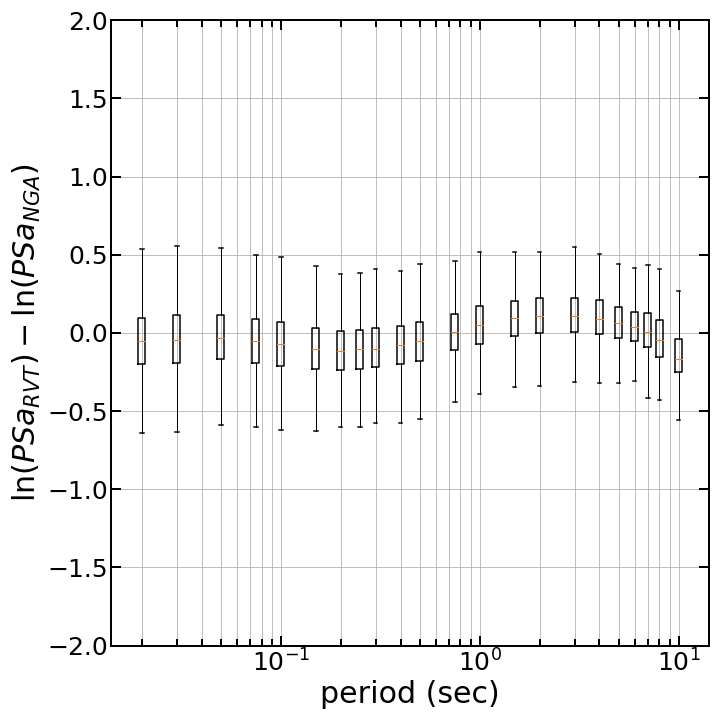}
    \end{subfigure}
    \caption{Residuals between the records' $PSa$ and the RVT $PSa$. V75 $PF$, $2~D_{a0.20-0.95}$ from KS06 as $D_{gm}$, and BT15 $D_{rms}$. (a) residuals for all $M$, (b) residuals for $M > 5$}
    \label{esup:fig:cmp_PF_BT15_dur_AS16_2da0.20-0.80}
\end{figure}

Figures \ref{esup:fig:cmp_PF_BT15_dur_KS06_da0.05-0.75} to \ref{esup:fig:cmp_PF_BT15_dur_AS16_2da0.20-0.80} show the residuals between the records' $PSa$ and the $PSa$ estimated with RVT. The RVT $PSa$ were estimated with $V75$ peak factors, the $D_{gm}$ estimated with a duration GMM, and $BT15$ for $D_{rms}$.
$D_{gm}$ is based on: KS06 $D_{a5-75}$ in Figure \ref{esup:fig:cmp_PF_BT15_dur_KS06_da0.05-0.75}, KS06 $D_{a5-95}$ in Figure \ref{esup:fig:cmp_PF_BT15_dur_KS06_da0.05-0.95}, KS06 $D_{v5-75}$ in Figure \ref{esup:fig:cmp_PF_BT15_dur_KS06_dv0.05-0.75}, the KS06 $D_{v5-95}$ in Figure \ref{esup:fig:cmp_PF_BT15_dur_KS06_dv0.05-0.95}, AS16 $D_{a5-75}$ in Figure \ref{esup:fig:cmp_PF_BT15_dur_AS16_da0.05-0.75}, AS16 $D_{a5-95}$ in Figure \ref{esup:fig:cmp_PF_BT15_dur_AS16_da0.05-0.95}, and AS16 $2~D_{a20-80}$ in Figure \ref{esup:fig:cmp_PF_BT15_dur_AS16_2da0.20-0.80}.

